# GAUDIN MODEL, BETHE ANSATZ AND CRITICAL LEVEL

BORIS FEIGIN, EDWARD FRENKEL, AND NIKOLAI RESHETIKHIN

1. INTRODUCTION.

Gaudin's model describes a completely integrable quantum spin chain. Originally [1] it was formulated as a spin model related to the Lie algebra $\mathfrak{sl}_2$. Later it was realized, cf. [2], Sect. 13.2.2 and [3], that one can associate such a model to any semi-simple complex Lie algebra $\mathfrak{g}$ and a solution of the corresponding classical Yang-Baxter equation [4, 5]. In this work we will focus on the models, corresponding to the rational solutions.

Denote by $V_\lambda$ the finite-dimensional irreducible representation of $\mathfrak{g}$ of dominant integral highest weight $\lambda$. Let $(\boldsymbol{\lambda}) := (\lambda_1, \ldots, \lambda_N)$ be a set of dominant integral weights of $\mathfrak{g}$. Consider the tensor product $V_{(\boldsymbol{\lambda})} := V_{\lambda_1} \otimes \ldots \otimes V_{\lambda_N}$ and associate with each factor $V_{\lambda_i}$ of this tensor product a complex number $z_i$. The hamiltonians of Gaudin's model are mutually commuting operators $\Xi_i = \Xi_i(z_1, \ldots, z_N), i = 1, \ldots, N$, acting on the space $V_{(\boldsymbol{\lambda})}$.

Denote by $\langle \cdot, \cdot \rangle$ the invariant scalar product on $\mathfrak{g}$, normalized as in [6]. Let $\{I_a\}, a = 1, \ldots, d = \dim \mathfrak{g}$, be a basis of $\mathfrak{g}$ and $\{I^a\}$ be the dual basis. For any element $A \in \mathfrak{g}$ denote by $A^{(i)}$ the operator $1 \otimes \ldots \otimes \underset{i}{A} \otimes \ldots \otimes 1$, which acts as $A$ on the $i$th factor of $V_{(\boldsymbol{\lambda})}$ and as the identity on all other factors. Then

$$(1.1) \qquad \Xi_i = \sum_{j \neq i} \sum_{a=1}^d \frac{I_a^{(i)} I^{a(j)}}{z_i - z_j}.$$

One of the main problems in Gaudin's model is to find the eigenvectors and the eigenvalues of these operators.

Bethe ansatz method is perhaps the most effective method for solving this problem for $\mathfrak{g} = \mathfrak{sl}_2$ [1]. As general references on Bethe ansatz, cf. [7, 8, 2, 9]. Applied to Gaudin's model, this method consists of the following. There is an obvious eigenvector in $V_{(\boldsymbol{\lambda})}$: the tensor product of the highest weight vectors of the $V_{\lambda_i}$'s. One constructs other eigenvectors by acting on this vector by certain elementary operators, depending on auxiliary parameters $w_1, \ldots, w_m$. The vectors obtained this way are called Bethe vectors. Such a vector is an eigenvector of the Gaudin hamiltonians, if a certain system of algebraic equations, involving the parameters $z_i$'s and $w_j$'s, and the highest weights $\lambda_i$'s, is satisfied. These equations are called Bethe ansatz equations.

One usually constructs eigenvectors in statistical models associated to a general simple Lie algebra $\mathfrak{g}$ by choosing a sequence of embeddings of Lie algebras of lower rank into $\mathfrak{g}$ and inductively solving diagonalization problems for these subalgebras [15, 16, 3]. This leads to a rather complicated combinatorial algorithm, which very much depends on the structure of $\mathfrak{g}$.

Another method was recently proposed by Babujian and Flume [37]. They consider as analogues of Bethe vectors, the rational $V_{(\lambda)}$-valued functions, which enter Schechtman-Varchenko solutions [30, 31] of the Knizhnik-Zamolodchikov (KZ) equation [27] associated to $\mathfrak{g}$, cf. also [36, 38]. Varchenko and one of us show elsewhere [39] that such a vector can be obtained as a quasi-classical asymptotics of a solution of the KZ equation. This establishes a remarkable connection between a model of statistical mechanics and correlation functions of a conformal field theory.

In this work we propose a new method of diagonalization of Gaudin's hamiltonians, which is based on Wakimoto modules over the affine algebra $\widehat{\mathfrak{g}}$ at the critical level and the concept of invariant functionals (correlation functions) on tensor products of $\widehat{\mathfrak{g}}$-modules. This will allow us to treat the diagonalization problem and the KZ equation on equal footing and thus explain the connection between them.

The critical level is the level $k = -h^{\vee}$, where $h^{\vee}$ is the dual Coxeter number of $\mathfrak{g}$. The peculiarity of this value of level is that it is only for $k = -h^{\vee}$ that the local completion $U_k(\widehat{\mathfrak{g}})_{\mathrm{loc}}$ of the universal enveloping algebra of $\widehat{\mathfrak{g}}$ contains central elements. The center of $U_{-h^{\vee}}(\widehat{\mathfrak{g}})_{\mathrm{loc}}$ was described in [10, 11] (cf. also [12, 13, 14]). Elements of the center act on the space of functionals on tensor products of representations of $\widehat{\mathfrak{g}}$ at the critical level, which are invariant with respect to a certain Lie algebra. One can interpret the hamiltonian $\Xi_i$ as the action of a quadratic central element from $U_{-h^{\vee}}(\widehat{\mathfrak{g}})_{\mathrm{loc}}$ on the $i$th factor of invariant functional. Other central elements give rise to other operators, which commute with each other and with Gaudin's hamiltonians. We call them higher Gaudin's hamiltonians.

Wakimoto modules [17, 18] at the critical level are special bosonic representations of $\widehat{\mathfrak{g}}$, which are essentially parametrized by functions on the circle with values in the dual space of the Cartan subalgebra $\mathfrak{h}$ of $\mathfrak{g}$. We will construct eigenvectors of Gaudin's hamiltonians by restricting certain invariant functionals on tensor products of Wakimoto modules. In conformal field theory language, those are certain bosonic correlation functions. Analogues of Bethe ansatz equations will naturally appear as Kac-Kazhdan type [19] equations on the existence of certain singular vectors in Wakimoto modules. The existence of these singular vectors will ensure that the vector constructed this way is an eigenvector of the Gaudin hamiltonians and their generalizations.

We will express the eigenvalues of these hamiltonians in terms of the Miura transformation and show that Bethe ansatz equations appear as certain analytic conditions on the eigenvalues. The formulas we obtain for the eigenvalues generalize Sklyanin's formula [20] for $\mathfrak{g} = \mathfrak{sl}_2$. The appearance of the Miura transformation in this context is not surprising, because the center of $U_{-h^\vee}(\widehat{\mathfrak{g}})_{\text{loc}}$ is isomorphic to the classical $\mathcal{W}$–algebra associated to $\mathfrak{g}^\vee$, the Langlands dual Lie algebra to $\mathfrak{g}$ [10].

The space of invariant functionals makes sense for an arbitrary level $k$. For non-critical values of $k$ it coincides with the space of genus 0 correlation functions (or conformal blocks) of the corresponding Wess-Zumino-Novikov-Witten (WZNW) model [21, 22, 23, 24, 25, 26]. This definition is equivalent to the more common definition of correlation functions as matrix elements of certain vertex operators, acting between representations of $\widehat{\mathfrak{g}}$. It is known that for non-critical $k$ these functions satisfy the KZ equation [27, 28, 29].

One can obtain solutions of the KZ equation by restricting certain invariant functionals on tensor products of Wakimoto modules, but those of *non-critical* level. The rational function, which enters such a solution [30, 31] (cf. also [32, 33, 34, 26]), then appears as a bosonic correlation function, cf. [35], which coincides with the formula for an eigevector, obtained from Wakimoto modules at the *critical* level. This gives an explanation of the connection between the eigenvectors of the Gaudin model and solutions of the KZ equation associated to $\mathfrak{g}$, which was observed in [36, 37, 38, 39].

Gaudin's hamiltonians can be obtained from the transfer-matrix of a quantum spin chain related to affine quantum groups $U_q(\widehat{\mathfrak{g}})$ in a certain limit, when $q \to 1$. A generalization of the results presented here to the case of $U_q(\widehat{\mathfrak{g}})$ will be published separately.

We also want to remark that our construction of eigenvectors naturally fits into a program of geometric Langlands correspondence proposed by Drinfeld. This correspondence relates equivalence classes of $G^\vee$–bundles over a complex algebraic curve $\mathcal{E}$ with flat connections, and $\mathcal{D}$–modules on the moduli space of $G$–bundles over $\mathcal{E}$.

The paper is organized as follows. In Sect. 2 we recall the definition of Gaudin's model and Bethe ansatz procedure in the case $\mathfrak{g} = \mathfrak{sl}_2$. Sect. 3 contains an interpretation of the model in terms of invariant functionals at the critical level. We show how singular vectors of imaginary weight give rise to Gaudin's hamiltonians and their generalizations. In Sect. 4 we recall the definition of Wakimoto modules and discuss the structure of the center of $U_{-h^\vee}(\widehat{\mathfrak{g}})_{\text{loc}}$. We use these results in Sect. 5 to construct eigenvectors of Gaudin's hamiltonians and to compute their eigenvalues. In Sect. 6 we derive Schechtman-Varchenko solutions of the KZ equation, using Wakimoto modules. We show that the rational function entering a solution coincides with the formula for an eigenvector. In the Appendix we prove some technical results concerning Wakimoto modules, which are used in the main text.

## 2. Gaudin's model.

Let $\mathfrak{g}$ be a simple Lie algebra over $\mathbb{C}$ of rank $l$ and dimension $d$, and let $U(\mathfrak{g})$ be its universal enveloping algebra. For a dominant integral weight $\lambda$ denote by $V_\lambda$ the irreducible representation of $\mathfrak{g}$ of highest weight $\lambda$. Denote by $\langle \cdot, \cdot \rangle$ the invariant scalar product on $\mathfrak{g}$, normalized as in [6]. Let $I_a, a = 1, \ldots, d$, be a root basis of $\mathfrak{g}$ and $I^a$ be the dual basis. Denote by $\Delta$ the quadratic Casimir operator from the center of $U(\mathfrak{g})$:

$$\Delta = \frac{1}{2} \sum_{a=1}^{d} I_a I^a.$$

The operator $\Delta$ acts on $V_\lambda$ by multiplication by a number, which we will denote by $\Delta(\lambda)$.

Now fix a positive integer $N$ and a set $(\boldsymbol{\lambda}) := (\lambda_1, \ldots, \lambda_N)$ of dominant highest weights. Denote by $V_{(\boldsymbol{\lambda})}$ the tensor product $V_{\lambda_1} \otimes \ldots \otimes V_{\lambda_N}$.

Let $u, z_1, \ldots, z_N$ be a set of distinct complex numbers. Introduce linear operator $S(u)$, which we will call the Gaudin hamiltonian, on $V_{(\boldsymbol{\lambda})}$ by the formula

$$(2.1) \qquad S(u) = \sum_{i=1}^{N} \frac{\Xi_i}{u - z_i} + \sum_{i=1}^{N} \frac{\Delta(\lambda_i)}{(u - z_i)^2},$$

where $\Xi_i$ is given by formula (1.1).

One can check directly [1, 3] that the operators $\Xi_1, \ldots, \Xi_N$ commute with each other. Therefore the operators $S(u)$ commute with each other for different values of $u$. (In the next section we will give a new proof of this fact.)

One of the main problems in Gaudin's model is to diagonalize operators $S(u)$. This is equivalent to simultaneous diagonalization of the operators $\Xi_1, \ldots, \Xi_N$.

For $\mathfrak{g} = \mathfrak{sl}_2$ this problem can be solved by algebraic Bethe ansatz [1, 2, 20].

Denote by $v_\lambda$ the highest weight vector of the module $V_\lambda$. The tensor product of the highest weight vectors

$$|0\rangle := v_{\lambda_1} \otimes \ldots \otimes v_{\lambda_N} \in V_{(\boldsymbol{\lambda})}$$

is an eigenvector of the operators $\Xi_i$. Indeed, if $I_a$ and $I^a$ are not elements of the Cartan subalgebra of $\mathfrak{g}$, then the operator $I_a^{(i)} I^{a(j)}$ acts by $0$ on $|0\rangle$. If they are, then this operator multiplies $|0\rangle$ by a certain number.

Therefore $|0\rangle$ is an eigenvector of $S(u)$.

The idea of the Bethe ansatz method is to produce new eigenvectors by applying certain elementary operators to the "vacuum" $|0\rangle$.

Let $\{E, H, F\}$ be the standard basis of $\mathfrak{sl}_2$. Introduce operators $F(w)$ on the space $V_{(\boldsymbol{\lambda})}$ by the formula

$$(2.2) \qquad F(w) = \sum_{i=1}^{N} \frac{F^{(i)}}{w - z_i},$$

where $F^{(i)}$ is the operator, which acts as the generator $F$ of $\mathfrak{sl}_2$ on the $i$th factor of $V_{(\boldsymbol{\lambda})}$ and as the identity on all other factors. Here $w$ is a complex number, which is not equal to $z_1, \ldots, z_N$.

Now consider *Bethe's vector*

$$|w_1, \ldots, w_m\rangle = F(w_1) \ldots F(w_m)|0\rangle \tag{2.3}$$

in $V_{(\lambda)}$.

Explicit computation [1] shows that

$$S(u)|w_1, \ldots, w_m\rangle = s_m(u)|w_1, \ldots, w_m\rangle + \sum_{j=1}^{m} \frac{f_j}{u - w_j}|w_1, \ldots, w_{j-1}, u, w_{j+1}, \ldots, w_m\rangle, \tag{2.4}$$

where $s_m(u)$ is a function in $u$, and

$$f_j = \sum_{i=1}^{N} \frac{\lambda_i}{w_j - z_i} - \sum_{s \neq j} \frac{2}{w_j - w_s}.$$

If all $f_j$'s vanish, then $|w_1, \ldots, w_m\rangle$ is an eigenvector of $S(u)$. The corresponding equations

$$\sum_{i=1}^{N} \frac{\lambda_i}{w_j - z_i} - \sum_{s \neq j} \frac{2}{w_j - w_s} = 0, \qquad j = 1, \ldots, m, \tag{2.5}$$

are called *Bethe ansatz equations*.

The eigenvalue of $S(u)$ on the vector $|w_1, \ldots, w_m\rangle$ is given by

$$s_m(u) = \frac{1}{4}\chi_m(u)^2 - \frac{1}{2}\partial_u \chi_m(u), \tag{2.6}$$

where

$$\chi_m(u) = \sum_{i=1}^{N} \frac{\lambda_i}{u - z_i} - \sum_{j=1}^{m} \frac{2}{u - w_j}. \tag{2.7}$$

There are several approaches to constructing eigenvectors of Gaudin's hamiltonians for general $\mathfrak{g}$.

One can construct eigenvectors by choosing a sequence of embeddings of Lie algebras of lower rank into $\mathfrak{g}$ and inductively solving the diagonalization problems for these subalgebras [15, 16, 3]. This leads to a rather complicated combinatorial algorithm, which very much depends on the type of $\mathfrak{g}$.

Another method was proposed by Babujian and Flume in [37]. The eigenvectors should be constructed by applying to the vacuum $|0\rangle$, the operators

$$F_j(w) = \sum_{i=1}^{N} \frac{F_j^{(i)}}{w - z_i}, \qquad j = 1, \ldots, l.$$

However, since such operators no longer commute with each other, one should also add some extra terms, which account for the commutators. The right formula can be extracted from solutions of the KZ equation [30, 31] (and in fact can be obtained as quasi-classical asymptotics of such solutions [39]):

$$(2.8) \quad |w_1^{i_1}, \ldots, w_m^{i_m}\rangle = \sum_{p=(I^1, \ldots, I^N)} \prod_{j=1}^{N} \frac{F_{i_1^j}^{(j)} F_{i_2^j}^{(j)} \ldots F_{i_{a_j}^j}^{(j)}}{(w_{i_1^j} - w_{i_2^j})(w_{i_2^j} - w_{i_3^j}) \ldots (w_{i_{a_j}^j} - z_j)} |0\rangle.$$

Here the summation is taken over all *ordered* partitions $I^1 \cup I^2 \cup \ldots \cup I^N$ of the set $\{1, \ldots, m\}$, where $I^j = \{i_1^j, i_2^j, \ldots, i_{a_j}^j\}$. Note that one can consider vector (2.8) as an element of the tensor product of Verma modules $M_{\lambda_1} \otimes \ldots \otimes M_{\lambda_N}$ with arbitrary highest weights $\lambda_1, \ldots, \lambda_N$.

It was claimed in [37] that the vector $|w_1^{i_1}, \ldots, w_m^{i_m}\rangle$ satisfies an analogue of the equation (2.4). Namely,

$$(2.9) \quad S(u)|w_1^{i_1}, \ldots, w_m^{i_m}\rangle = s_{i_1,\ldots,i_m}(u)|w_1^{i_1}, \ldots, w_m^{i_m}\rangle + \sum_{j=1}^{m} \frac{f_j^{i_j}}{u - w_j} X_j,$$

where $s_{i_1,\ldots,i_m}(u)$ is a rational function in $u$,

$$(2.10) \quad f_j^{i_j} = \sum_{i=1}^{N} \frac{(\lambda_i, \alpha_{i_j})}{w_j - z_i} - \sum_{s \neq j} \frac{(\alpha_{i_s}, \alpha_{i_j})}{w_j - w_s},$$

and $X_j$ is a certain vector in $V_{(\lambda)}$. Thus, if the equations $f_j^{i_j} = 0$ are satisfied for all $j = 1, \ldots, m$, then vector $|w_1^{i_1}, \ldots, w_m^{i_m}\rangle$ is an eigenvector of Gaudin's hamiltonians. It is natural to call these equations Bethe equations and the vector $|w_1^{i_1}, \ldots, w_m^{i_m}\rangle$ a Bethe vector.

The direct proof of formula (2.9) seems rather complicated, and it is not clear how to generalize it to higher Gaudin's hamiltonians, which we introduce in the next section.

In this work we propose an alternative approach to the diagonalization of Gaudin's hamiltonians, which is based on a concept of invariant functionals (correlation functions) on tensor products of representations of affine algebras at the critical level.

In the next section we will interpret Gaudin's model in terms of such functionals. We will show that Gaudin's hamiltonians are members of a family of commuting linear operators on the invariant functionals, which come from singular vectors of imaginary weight in the vacuum representation of $\widehat{\mathfrak{g}}$ at the critical level.

In Sect. 5 we will construct eigenvectors of Gaudin's hamiltonians by restricting certain invariant functionals on tensor products of Wakimoto modules. This construction will allow us to prove that if Bethe equations are satisfied, then vector $|w_1^{i_1}, \ldots, w_m^{i_m}\rangle$ is an eigenvector of the operators $S(u)$ and their generalizations, defined in the next section. We will also compute the eigenvalues of these operators. This construction will then be used in Sect. 6 to give an explanation of the appearance of Bethe vectors in solutions of the KZ equation.

## 3. Correlation functions.

Let $\widehat{\mathfrak{g}}$ be the affine algebra, corresponding to $\mathfrak{g}$. It is the extension of the Lie algebra $\mathfrak{g} \otimes \mathbb{C}((t))$ by one-dimensional center $\mathbb{C}K$. Denote by $\widehat{\mathfrak{g}}_+$ the Lie subalgebra $\mathfrak{g} \otimes \mathbb{C}[[t]] \oplus \mathbb{C}K$ of $\widehat{\mathfrak{g}}$. For any finite-dimensional representation $V_\lambda$ of $\mathfrak{g}$ denote by $V_\lambda^k$ the representation of $\widehat{\mathfrak{g}}_+$, on which $\mathfrak{g} \otimes t\mathbb{C}[[t]]$ acts trivially and $K$ acts by multiplication by $k \in \mathbb{C}$. Denote by $\mathbb{V}_\lambda^k$ the induced representation of $\widehat{\mathfrak{g}}$ of level $k$:

$$\mathbb{V}_\lambda^k = U(\widehat{\mathfrak{g}}) \otimes_{U(\widehat{\mathfrak{g}}_+)} V_\lambda^k.$$

Consider the projective line $\mathbb{CP}^1$ with a global coordinate $t$ and $N$ distinct finite points $z_1, \ldots, z_N \in \mathbb{CP}^1$. In the neighbourhood of each point $z_i$ we have the local coordinate $t - z_i$. Denote $\widetilde{\mathfrak{g}}(z_i) = \mathfrak{g} \otimes \mathbb{C}((t - z_i))$. Let $\widehat{\mathfrak{g}}_N$ be the extension of the Lie algebra $\oplus_{i=1}^N \widetilde{\mathfrak{g}}(z_i)$ by one-dimensional center $\mathbb{C}K$, such that its restriction to any summand $\widetilde{\mathfrak{g}}(z_i)$ coincides with the standard extension. For $\mathbf{f} = (f_1(t-z_1), \ldots, f_N(t-z_N))$ and $\mathbf{g} = (g_1(t-z_1), \ldots, g_N(t-z_N))$ from $\oplus_{i=1}^N \widetilde{\mathfrak{g}}(z_i)$ the cocycle $\omega_N$, defining this central extension, is given by

$$(3.1) \qquad \omega_N(\mathbf{f}, \mathbf{g}) = K \cdot \sum_{i=1}^N \frac{1}{2\pi i} \int_i \langle f_i, dg_i \rangle,$$

where $\int_i$ stands for the integral over a small contour around the point $z_i$.

The Lie algebra $\widehat{\mathfrak{g}}_N$ naturally acts on the tensor product $\mathbb{V}_{(\boldsymbol{\lambda})}^k = \mathbb{V}_{\lambda_1}^k \otimes \ldots \otimes \mathbb{V}_{\lambda_N}^k$, in particular, $K$ acts by multiplication by $k$. We will say that we have *assigned* modules $\mathbb{V}_{\lambda_1}^k, \ldots, \mathbb{V}_{\lambda_N}^k$ to the points $z_1, \ldots, z_N$.

Let $\mathfrak{g}_\mathbf{z} := \mathfrak{g}_{z_1, \ldots, z_N}$ be the Lie algebra of $\mathfrak{g}$–valued regular functions on $\mathbb{CP}^1 \backslash \{z_1, \ldots, z_N\}$ (i.e. rational functions on $\mathbb{CP}^1$, which may have poles only at the points $z_1, \ldots, z_N$), which vanish at infinity. Clearly, such a function can be expanded into a Laurent power series in the local coordinate $t - z_i$ at each point $z_i$. This gives an element of $\widetilde{\mathfrak{g}}(z_i)$. Thus, we obtain an embedding of $\mathfrak{g}_\mathbf{z}$ into the direct sum $\oplus_{i=1}^N \widetilde{\mathfrak{g}}(z_i)$. The restriction of the central extension to the image of this embedding is trivial. Indeed, according to formula (3.1), this restriction is given by the sum of all residues of a certain one-form on $\mathbb{CP}^1$; hence it should vanish. Therefore we can lift the embedding $\mathfrak{g}_\mathbf{z} \to \oplus_{i=1}^N \widetilde{\mathfrak{g}}(z_i)$ to an embedding $\mathfrak{g}_\mathbf{z} \to \widehat{\mathfrak{g}}_N$.

Denote by $H_{(\boldsymbol{\lambda})}^k$ the space of linear functionals on $\mathbb{V}_{(\boldsymbol{\lambda})}^k$, which are invariant with respect to the action of the Lie algebra $\mathfrak{g}_\mathbf{z}$. Such a functional is a linear map $\mu : \mathbb{V}_{(\boldsymbol{\lambda})}^k \to \mathbb{C}$, which satisfies

$$(3.2) \qquad \mu(g \cdot y) = 0, \qquad \forall g \in \mathfrak{g}_\mathbf{z}, y \in \mathbb{V}_{(\boldsymbol{\lambda})}.$$

By construction, we have a canonical embedding of a finite-dimensional representation $V_\lambda$ into the module $\mathbb{V}_\lambda^k$:

$$x \in V_\lambda \to 1 \otimes x \in \mathbb{V}_\lambda,$$

which commutes with the action of $\mathfrak{g}$ on both spaces (recall that $\mathfrak{g}$ is embedded into $\widehat{\mathfrak{g}}$ as the constant subalgebra). Thus we have an embedding of $V_{(\boldsymbol{\lambda})} = V_{\lambda_1} \otimes \ldots \otimes V_{\lambda_N}$ into $\mathbb{V}_{(\boldsymbol{\lambda})}^k$. We will keep the same notation $V_{(\boldsymbol{\lambda})}$ for the image of this embedding.

The restriction of any invariant functional $\mu \in H_{(\boldsymbol{\lambda})}^k$ to the subspace $V_{(\boldsymbol{\lambda})} \subset \mathbb{V}_{(\boldsymbol{\lambda})}^k$ defines a linear functional on $V_{(\boldsymbol{\lambda})}$. Thus we obtain a map $\epsilon_N^k : H_{(\boldsymbol{\lambda})}^k \to V_{(\boldsymbol{\lambda})}^*$.

**Lemma 1.** *The map $\epsilon_N^k : H_{(\boldsymbol{\lambda})}^k \to V_{(\boldsymbol{\lambda})}^*$ is an isomorphism.*

*Proof.* Denote $\widetilde{\mathfrak{g}}_+(z_i) := \mathfrak{g} \otimes \mathbb{C}[[t-z_i]] \subset \widetilde{\mathfrak{g}}(z_i)$. We have a direct sum decomposition

$$\oplus_{i=1}^N \widetilde{\mathfrak{g}}(z_i) = \left(\oplus_{i=1}^N \widetilde{\mathfrak{g}}_+(z_i)\right) \oplus \mathfrak{g}_{\mathbf{z}}.$$

Indeed, for $g(t) \in \mathfrak{g} \otimes \mathbb{C}((t))$ denote by $g^+(t)$ and $g^-(t)$ its regular and singular parts, respectively. Any element $\mathbf{f} = (f_1(t-z_1), \ldots, f_N(t-z_N))$ of $\oplus_{i=1}^N \widetilde{\mathfrak{g}}(z_i)$ can be uniquely represented as the sum of

$$f^- = \sum_{i=1}^N f_i^-(t-z_i) \in \mathfrak{g}_{\mathbf{z}}$$

and

$$(f_1^+(t-z_1) - (f^-)_1(t-z_1), \ldots, f_N^+(t-z_N) - (f^-)_N(t-z_N)) \in \oplus_{i=1}^N \widetilde{\mathfrak{g}}_+(z_i),$$

where $(f^-)_i(t-z_i)$ denotes the regular part of the expansion of the function $f^-$ at $z_i$.

The $\widehat{\mathfrak{g}}_N$–module $\mathbb{V}_{(\boldsymbol{\lambda})}^k$ is induced from the module $V_{(\boldsymbol{\lambda})}$ over $\oplus_{i=1}^N \widetilde{\mathfrak{g}}_+(z_i) \oplus \mathbb{C}K$, on which $\oplus_{i=1}^N (t-z_i) \cdot \widetilde{\mathfrak{g}}_+(z_i)$ acts trivially and $K$ acts by multiplication by $k$. Therefore, as a $\mathfrak{g}_{\mathbf{z}}$–module, $\mathbb{V}_{(\boldsymbol{\lambda})}^k$ is isomorphic to the free module generated by $V_{(\boldsymbol{\lambda})}$. Hence the space of $\mathfrak{g}_{\mathbf{z}}$–invariant functionals on $\mathbb{V}_{(\boldsymbol{\lambda})}^k$ is isomorphic to $V_{(\boldsymbol{\lambda})}^*$. □

Lemma 1 shows that an invariant functional is uniquely defined by its restriction to the subspace $V_{(\boldsymbol{\lambda})} \subset \mathbb{V}_{(\boldsymbol{\lambda})}^k$. Thus, for any $\eta \in V_{(\boldsymbol{\lambda})}^*$ there exists a $\mathfrak{g}_{\mathbf{z}}$–invariant functional $\widetilde{\eta} := (\epsilon_N^k)^{-1} \eta : \mathbb{V}_{(\boldsymbol{\lambda})}^k \to \mathbb{C}$, such that the restriction of $\widetilde{\eta}$ to $V_{(\boldsymbol{\lambda})} \subset \mathbb{V}_{(\boldsymbol{\lambda})}^k$ coincides with $\eta$.

Consider the Lie algebra $\mathfrak{g}_{\mathbf{z}}^0 = \mathfrak{g}_{\mathbf{z}} \oplus \mathfrak{g} \subset \widehat{\mathfrak{g}}_N$, where $\mathfrak{g}$ is the constant diagonal subalgebra of $\widehat{\mathfrak{g}}_N$. Note that $\mathfrak{g}$ does not lie in $\mathfrak{g}_{\mathbf{z}}$, because by definition elements of $\mathfrak{g}_{\mathbf{z}}$ must vanish at infinity. One can consider the space of $\mathfrak{g}_{\mathbf{z}}^0$–invariant functionals on $\mathbb{V}_{(\boldsymbol{\lambda})}^k$. This space is isomorphic to the space of $\mathfrak{g}$–invariants of $V_{(\boldsymbol{\lambda})}$. In conformal field theory a $\mathfrak{g}_{\mathbf{z}}^0$–invariant functional $\mu$ on $\mathbb{V}_{(\boldsymbol{\lambda})}^k$ is called a *correlation function* (more precisely, conformal block), and the equation (3.2) is called *Ward's identity*.

*Remark* 1. Sometimes it is convenient to consider instead of the space of $\mathfrak{g}_{\mathbf{z}}$–invariant functionals on $\mathbb{V}_{(\boldsymbol{\lambda})}^k$, the space of coinvariants of $\mathbb{V}_{(\boldsymbol{\lambda})}$ with respect to $\mathfrak{g}_{\mathbf{z}}$. These spaces are dual to each other. For a detailed study of the functor of coinvariants, cf. [25]. □

Let $M$ be a $\mathfrak{g}$–module, and let $M^*$ be the restricted dual linear space to $M$. We can define a structure of $\mathfrak{g}$–module on $M^*$ as follows:

(3.3) $$[g \cdot f](m) = f(\imath(g) \cdot Y), \qquad f \in M^*, g \in \mathfrak{g}, m \in M,$$

where $\imath$ stands for the Cartan anti-involution on $\mathfrak{g}$. It is defined on the generators of $\mathfrak{g}$ as follows:
$$\imath(E_i) = F_i, \qquad \imath(F_i) = E_i, \qquad \imath(H_i) = H_i.$$

This gives a structure of $\mathfrak{g}$–module on $M^*$, which is called contragradient to $M$. It is known that if the module $M$ belongs to the category $\mathcal{O}$ of $\mathfrak{g}$–modules, so does $M^*$. In particular, the module contragradient to an irreducible representation $V_\lambda$ with highest weight $\lambda$ is isomorphic to itself. Hence as a contragradient module, $V_{(\boldsymbol{\lambda})}^*$ is isomorphic to $V_{(\boldsymbol{\lambda})}$. Therefore $H_{(\boldsymbol{\lambda})}^k \simeq V_{(\boldsymbol{\lambda})}$.

Let $\mathbb{V}_0^k$ be the representation of $\widehat{\mathfrak{g}}$, which corresponds to the one-dimensional trivial $\mathfrak{g}$–module $V_0$. We will call it the *vacuum module*. Denote by $v_0$ the generating vector of $\mathbb{V}_0^k$. We assign the vacuum module to a finite point $u \in \mathbb{CP}^1$, which is different from $z_1, \ldots, z_N$. Denote by $H_{(\boldsymbol{\lambda},0)}^k$ the space of $\mathfrak{g}_{\mathbf{z},u}$–invariant functionals on $\mathbb{V}_{(\boldsymbol{\lambda})}^k \otimes \mathbb{V}_0^k$ with respect to the Lie algebra $\mathfrak{g}_{\mathbf{z},u}$. Lemma 1 tells us that $H_{(\boldsymbol{\lambda},0)}^k \simeq V_{(\boldsymbol{\lambda})}^*$.

Let $X$ be a vector in $\mathbb{V}_0^k$. For any $\eta \in V_{(\boldsymbol{\lambda})}^*$ consider the corresponding $\mathfrak{g}_{\mathbf{z},u}$–invariant functional $\widetilde{\eta} \in H_{(\boldsymbol{\lambda},0)}^k$. Its restriction $\widetilde{\eta}(\cdot, X)$ to the subspace $V_{(\boldsymbol{\lambda})} \otimes X \subset \mathbb{V}_{(\boldsymbol{\lambda},0)}$ defines another linear functional, $\eta'(\cdot)$, on $V_{(\boldsymbol{\lambda})}$. Thus, we obtain a linear operator depending on $u$, $X(u) : V_{(\boldsymbol{\lambda})}^* \to V_{(\boldsymbol{\lambda})}^*$, which sends $\eta$ to $\eta'$. Since $V_{(\boldsymbol{\lambda})}^* \simeq V_{(\boldsymbol{\lambda})}$, we can consider $X(u)$ as an operator acting on $V_{(\boldsymbol{\lambda})}$.

For $A \in \mathfrak{g}$ and $m \in \mathbb{Z}$, denote by $A(m)$ the element $A \otimes t^m \in \widehat{\mathfrak{g}}$. Now introduce the following vector in $\mathbb{V}_0^k$:

(3.4) $$S = \frac{1}{2} \sum_{a=1}^d I_a(-1) I^a(-1) v_0.$$

This vector defines a linear operator
$$S(u) : V_{(\boldsymbol{\lambda})} \to V_{(\boldsymbol{\lambda})}.$$

**Proposition 1.** *The operator $S(u)$ coincides with the Gaudin hamiltonian.*

*Proof.* For any $A \in \mathfrak{g}$ consider the element
$$g = \frac{A}{(t-u)^n} \in \mathfrak{g}_{\mathbf{z},u}.$$

The expansion of $g$ at $z_i$ is equal to
$$\frac{A}{(z_i - u)^n} \cdot \frac{1}{\left(1 - \frac{t-z_i}{u-z_i}\right)^n}.$$

Therefore the action of $g$ on the $i$th factor of the tensor product of $\mathbb{V}^k_{(\boldsymbol{\lambda},0)}$ is given by

$$-\frac{1}{(n-1)!}\frac{\partial^{n-1}}{\partial u^{n-1}}\sum_{m=0}^{\infty}\frac{A(m)^{(i)}}{(u-z_i)^{m+1}}.$$

Thus from (3.2) we obtain for any $Y \in \mathbb{V}^k_{(\boldsymbol{\lambda})}$

$$(3.5) \quad \widetilde{\eta}(Y, A(-n)X) = \frac{1}{(n-1)!}\frac{\partial^{n-1}}{\partial u^{n-1}}\widetilde{\eta}\left(\sum_{i=1}^{N}\sum_{m=0}^{\infty}\frac{A(m)^{(i)}}{(u-z_i)^{m+1}}\cdot Y, X\right).$$

This identity allows to "swap" the operator $A(-n)$ from the module, assigned to the point $u$, to the modules, assigned to the points $z_1, \ldots, z_N$.

Put $n = 1$ in (3.5). Since any element of $\mathfrak{g} \otimes (t - z_i)\mathbb{C}[[t - z_i]]$ acts trivially on the subspace $V_{(\boldsymbol{\lambda})}$ of $\mathbb{V}^k_{(\boldsymbol{\lambda})}$, we obtain for any $\omega \in V_{(\boldsymbol{\lambda})}$:

$$(3.6) \quad \widetilde{\eta}(\omega, A(-1)X) = \widetilde{\eta}\left(\sum_{i=1}^{N}\frac{A^{(i)}\cdot\omega}{u-z_i}, X\right), \quad X \in \mathbb{V}^k_0.$$

Applying this identity twice to the element $S \in \mathbb{V}^k_0$, we obtain:

$$\widetilde{\eta}(\omega, S) = \widetilde{\eta}\left(\frac{1}{2}\sum_{i=1}^{N}\sum_{a=1}^{d}\frac{I^{a(i)}}{u-z_i}\sum_{j=1}^{N}\sum_{a=1}^{d}\frac{I_a^{(j)}}{u-z_j}\cdot\omega, v_0\right) =$$

$$= \left[\frac{1}{2}\sum_{j=1}^{N}\sum_{a=1}^{d}\frac{\imath(I_a)^{(j)}}{u-z_j}\sum_{i=1}^{N}\sum_{a=1}^{d}\frac{\imath(I^a)^{(i)}}{u-z_i}\cdot\eta\right](\omega),$$

by formula (3.3). Hence

$$S(u) = \frac{1}{2}\sum_{i=1}^{N}\sum_{a=1}^{d}\frac{I^{a(i)}I_a^{(i)}}{(u-z_i)^2} + \sum_{i<j}\sum_{a=1}^{d}\frac{I_a^{(i)}I^{a(j)}}{(u-z_i)(u-z_j)}.$$

But this coincides with formula (2.1), because

$$\frac{1}{(u-z_i)(u-z_j)} = \frac{1}{z_i-z_j}\left(\frac{1}{u-z_i}-\frac{1}{u-z_j}\right).$$

□

We can now prove that the operators $S(u)$ commute for different values of $u$. In fact, we will prove a more general result.

Let us specialize to the value $k = -h^{\vee}$, where $h^{\vee}$ is the dual Coxeter number. This value of level is called critical. For simplicity in what follows we will omit the superscript $k$, when $k = -h^{\vee}$. Thus, we will write $\mathbb{V}_0$ instead of $\mathbb{V}_0^{-h^{\vee}}$, etc.

This value is special, because it is only at the critical level that $S \in \mathbb{V}^k_0$ is a singular vector of imaginary weight. This means that $A \cdot X = 0$ for any $A \in \mathfrak{g} \otimes \mathbb{C}[t]$.

**Proposition 2.** *Let $Z_1$ and $Z_2$ be singular vectors of imaginary weight in $\mathbb{V}_0$. Then for any pair of complex numbers $u$ and $v$ the corresponding linear operators $Z_1(u)$ and $Z_2(v)$ on $V_{(\boldsymbol{\lambda})}$ commute. In particular, for Gaudin's hamiltonians we have:*

$$[S(u), S(v)] = 0.$$

*Proof.* Consider a generalization of the previous construction, with two auxiliary points, $u$ and $v$. We can assign to each of these points the module $\mathbb{V}_0$. The space $H_{(\boldsymbol{\lambda},0,0)}$ is again isomorphic to $V_{(\boldsymbol{\lambda})}$. Therefore we can define an action of a pair of elements, $X_1, X_2 \in \mathbb{V}_0$, on $V_{(\boldsymbol{\lambda})}$. Namely, for any $\eta \in V_{(\boldsymbol{\lambda})}$ we consider the corresponding $\mathfrak{g}_{\mathbf{z},u,v}$–invariant functional $\widetilde{\eta} \in H_{(\boldsymbol{\lambda},0,0)}$. Its restriction $\widetilde{\eta}(\cdot, X_1, X_2)$ to the subspace $V_{(\boldsymbol{\lambda})} \otimes X_1 \otimes X_2 \subset \mathbb{V}_{(\boldsymbol{\lambda},0,0)}$ defines another linear functional, $\eta'(\cdot)$, on $V_{(\boldsymbol{\lambda})}$. Thus, we obtain a linear operator depending on $u$ and $v$, $(X_1, X_2)(u,v) : V_{(\boldsymbol{\lambda})}^* \to V_{(\boldsymbol{\lambda})}^*$, which sends $\eta$ to $\eta'$.

We will show now that if $X_1$ and $X_2$ are singular vectors of imaginary weight, then $(X_1, X_2)(u,v) = X_1(u)X_2(v)$ and $(X_1, X_2)(u,v) = X_2(v)X_1(u)$. This will prove that $[X_1(u), X_2(v)] = 0$.

Indeed, $X_1$ can be written as a linear combination of monomials $A_1(-n_1)\ldots A_m(-n_m)v_0$, where $A_j \in \mathfrak{g}, n_j > 0$. We can apply Ward's identity (3.6) to each summand and inductively "swap" operators $A_j(-n_j)$ to the modules assigned to the points $z_1, \ldots, z_N$ and $v$. This amounts to acting on them by

$$\frac{1}{(n_j-1)!}\sum_{i=1}^{N}\frac{A^{(i)}}{(v-z_i)^{n_j}} + \frac{1}{(n_j-1)!}\frac{\partial^{n_j-1}}{\partial u^{n_j-1}}\left(\sum_{m=0}^{\infty}\frac{A_j(m)^{(v)}}{(u-v)^{m+1}}\right).$$

But the last term of such a sum vanishes if we apply it to $X_2$, by definition of singular vector of imaginary weight. Therefore this only affects the modules at the points $z_i$, just as if we had $v_0$ at the point $v$ instead of $X_2$. After that we can "swap" $X_2$.

We obtain:

$$\widetilde{\eta}(\omega, X_1, X_2) = [X_1(u)X_2(v) \cdot \eta](\omega).$$

Hence $(X_1, X_2)(u,v) = X_1(u)X_2(v)$. In the same way we can show that $(X_1, X_2)(u,v) = X_2(v)X_1(u)$, and Proposition follows. $\square$

A description of the space $\mathcal{Z}(\widehat{\mathfrak{g}})$ of singular vectors of imaginary weight in $\mathbb{V}_0$ is known [11, 10]. It is related to the structure of the center of the local completion $U_{-h^\vee}(\widehat{\mathfrak{g}})_{\mathrm{loc}}$ of the universal enveloping algebra of $\widehat{\mathfrak{g}}$ at the critical level [11, 10]. Let us recall this description.

Recall that as a module over $\widehat{\mathfrak{g}}_- = \mathfrak{g} \otimes t^{-1}\mathbb{C}[t^{-1}]$, $\mathbb{V}_0$ is isomorphic to its universal enveloping algebra $U(\widehat{\mathfrak{g}}_-)$. Introduce a $\mathbb{Z}$-grading on $U(\widehat{\mathfrak{g}}_-)$ and on $\mathbb{V}_0$ by putting $\deg A(m) = -m, \deg v_0 = 0$. There is an operator of derivative $\partial = L_{-1}$ of degree 1 on the module $\mathbb{V}_0$ and hence on $U(\widehat{\mathfrak{g}}_-)$, such that $[\partial, A(m)] = -mA(m-1)$, and $\partial \cdot v_0 = 0$.

Denote by $d_1, \ldots, d_l$ the exponents of $\mathfrak{g}$.

**Proposition 3.** *The space $\mathcal{Z}(\widehat{\mathfrak{g}})$ of singular vectors of imaginary weight of $\mathbb{V}_0$ coincides with the polynomial algebra in $\partial^n S_i, i = 1, \ldots, l, n \geq 0$ (applied to vector $v_0$), where $S_i, i = 1, \ldots, l$, are mutually commuting elements of $U(\widehat{\mathfrak{g}}_-)$ of degrees $d_i + 1, i = 1, \ldots, l$.*

A more precise description of the center will be given in the next section. Note that $S_1$ is equal to $S$.

Proposition 3 and Proposition 2 show that $S_1(u), \ldots, S_l(u)$ constitute a family of mutually commuting linear operators on the space $V_{(\lambda)}$. We call $S_i(u), i > 1$, higher Gaudin's hamiltonians. It would be interesting to find explicit formulas for them.

*Remark 2.* For $Z \in \mathcal{Z}(\widehat{\mathfrak{g}})$ denote $\Xi_i^m(Z) = \text{Res}_{u=z_i}(u-z_i)^m Z(u)$. These operators generalize the operators $\Xi_i = \Xi_i^0(S)$, given by (1.1). They all mutually commute. Alternatively these operators can be defined as follows.

Since $\mathbb{V}_0$ is a vertex algebra [48], we can associate a local current $X(z)$ to any $X \in \mathbb{V}_0$. Denote by $X_{-m-1}$ the $(-m-1)$st Fourier component of this current. This is a central element of $U_{-h^\vee}(\widehat{\mathfrak{g}})_{\text{loc}}$ [10]. Therefore it acts on the $i$th component of the tensor product $\mathbb{V}_{\lambda_1} \otimes \ldots \otimes \mathbb{V}_{\lambda_N}$. The dual operator gives rise to an operator on the space of $\mathfrak{g}_z$-invariant functionals, $H_{(\lambda)} \simeq V_{(\lambda)}$. Using the generalized Ward identity from Sect. 5, we can show that the corresponding operator on $V_{(\lambda)}$ coincides with $\Xi_i^m(Z)$. □

Note that we can define the space of $\mathfrak{g}_z$-invariant functionals in a more general situation. To any $\mathfrak{g}$-module $M$ we can associate the induced $\widehat{\mathfrak{g}}$-module of level $k$,

$$\mathbb{M} = U(\widehat{\mathfrak{g}}) \otimes_{U(\widehat{\mathfrak{g}}_+)} M.$$

Fix $N$ representations $M_1, \ldots, M_N$ of $\mathfrak{g}$, and let $\mathbb{M}_1, \ldots, \mathbb{M}_N$ be the induced $\widehat{\mathfrak{g}}$-modules. We can define the space of $\mathfrak{g}_z$-invariant functionals on the tensor product $\mathbb{M}_1 \otimes \ldots \otimes \mathbb{M}_N$. This space is isomorphic to $M_1^* \otimes \ldots \otimes M_N^*$, where $M_i^*$ denotes the module contragradient to $M_i$. In the same way as above we can associate to a singular vector of imaginary weight $Z \in \mathbb{V}_0$, a family of commuting linear operators on this space.

It is clear from construction that the operators $Z(u)$ can be viewed as elements of the $N$th tensor power $U(\mathfrak{g})^{\otimes N}$ of the universal enveloping algebra of $\mathfrak{g}$, depending on $z_1, \ldots, z_N$, and $u$. Therefore they automatically act on any $N$-fold tensor product of $\mathfrak{g}$-modules.

Denote by $M_\lambda$ the Verma module over with highest weight $\lambda \in \mathfrak{h}^*$ over $\mathfrak{g}$, and by $M_\lambda^*$ the contragradient module.

From now on we will choose as our modules, $M_{\lambda_i}^*$ with arbitrary $\lambda_i \in \mathfrak{h}^*, i = 1, \ldots, l$. Then the space of $\mathfrak{g}_z$-invariant functionals is isomorphic to the tensor product of Verma modules, $\otimes_{i=1}^N M_{\lambda_i}$.

If $\lambda_i$'s are integral dominant weights, then there is a surjective $\mathfrak{g}^{\oplus N}$–homomorphism $\otimes_{i=1}^{N} M_{\lambda_i} \to \otimes_{i=1}^{N} V_{\lambda_i}$. The operators $Z(u), Z \in \mathcal{Z}(\widehat{\mathfrak{g}})$, act on both spaces and commute with this homomorphism. Thus, the projection of any eigenvector of $Z(u)$ from $\otimes_{i=1}^{N} M_{\lambda_i}$ on $\otimes_{i=1}^{N} V_{\lambda_i}$ is again an eigenvector.

Our construction of eigenvectors of Gaudin's hamiltonians in the tensor product $M_{\lambda_1} \otimes \ldots \otimes M_{\lambda_N}$ will be based on Wakimoto modules.

## 4. WAKIMOTO MODULES AT THE CRITICAL LEVEL.

In this section we briefly describe important facts about Wakimoto modules [17, 18, 40] (cf. also [41]).

Let us first introduce the Heisenberg Lie algebra $\Gamma(\mathfrak{g})$. It has generators $a_\alpha(m)$, $a_\alpha^*(m), m \in \mathbb{Z}, \alpha \in \Delta_+$, where $\Delta_+$ is the set of positive roots of $\mathfrak{g}$, and central element $\mathbf{1}$. They satisfy the relations:

$$[a_\alpha(n), a_\beta^*(m)] = \delta_{\alpha,\beta}\delta_{n,-m}\mathbf{1}, \quad [a_\alpha(n), a_\beta(m)] = 0, \quad [a_\alpha^*(n), a_\beta^*(m)] = 0.$$

Now let $M$ be the irreducible representation of $\Gamma(\mathfrak{g})$, which is generated from the *vacuum vector* $v$, satisfying

(4.1) $$a_\alpha(m)v = 0, m \geq 0, \qquad a_\alpha^*(m)v = 0, m > 0,$$

for all $\alpha \in \Delta_+$, and $\mathbf{1}v = v$.

Let $\widetilde{\mathfrak{h}}$ be the commutative Lie algebra $\mathfrak{h} \otimes \mathbb{C}((z))$. It has a linear basis $h_i(n) = H_i \otimes t^n, i = 1, \ldots, l, n \in \mathbb{Z}$. Any one-form $\chi(z)dz \in \mathfrak{h}^* \otimes \mathbb{C}((z))dz$ defines a one-dimensional representation $\sigma_{\chi(z)}$ of $\widetilde{\mathfrak{h}}$:

$$\sigma_{\chi(z)}[f(z)] = \frac{1}{2\pi i} \int \chi(z)[f(z)]dz, \qquad f(z) \in \widetilde{\mathfrak{h}},$$

where the integral is taken around the origin.

Wakimoto modules constitute a family of representations of the affine algebra $\widehat{\mathfrak{g}}$ of the critical level on the space $M \otimes \sigma_{\chi(z)}$.

In order to define these representations, we should construct a homomorphism $\rho$ from $\widehat{\mathfrak{g}}$ to the local completion of $U_1(\Gamma(\mathfrak{g})) \otimes U(\widetilde{\mathfrak{h}})$. Here $U_1(\Gamma(\mathfrak{g}))$ stands for the Heisenberg algebra, which is the quotient of the universal enveloping algebra of $\Gamma(\mathfrak{g})$ by the ideal generated by $(\mathbf{1} - 1)$. We recall the construction of the homomorphism $\rho$.

First we have to choose coordinates on the big cell $U$ of the flag manifold $B_-\backslash G$ of $\mathfrak{g}$. This big cell is isomorphic to the nilpotent subgroup $N_+$ of the Lie group $G$ of $\mathfrak{g}$. Since $N_+$ is isomorphic to its Lie algebra $\mathfrak{n}_+$ via the exponential map, $U$ is isomorphic to a linear space with coordinates $x_\alpha, \alpha \in \Delta_+$. We assign to the coordinate $x_\alpha$ the degree $\alpha$. The right action of $G$ on the flag manifold gives an embedding of the Lie algebra $\mathfrak{g}$ into the Lie algebra of vector fields on $U$. This embedding can be lifted to a family of embeddings into the Lie algebra of the first order differential operators, depending on $\chi \in \mathfrak{h}^*$. Denote these embeddings by $\bar{\rho}_\chi$. Explicitly, we have for the generators $E_i, H_i, F_i, i = 1, \ldots, l$, of $\mathfrak{g}$:

$$\bar{\rho}_\chi(E_i) = \frac{\partial}{\partial x_{\alpha_i}} + \sum_{\beta \in \Delta_+} P^i_\beta \frac{\partial}{\partial x_\beta},$$

where $P^i_\beta$ is a certain polynomial in $x_\alpha$ of degree $\beta - \alpha_i$;

$$\bar{\rho}_\chi(H_i) = -\sum_{\beta \in \Delta_+} \beta(H_i) x_\beta \frac{\partial}{\partial x_\beta} + \chi(H_i),$$

and

$$\bar{\rho}_\chi(F_i) = \sum_{\beta \in \Delta_+} Q^i_\beta \frac{\partial}{\partial x_\beta} + \chi(H_i) x_{\alpha_i},$$

where $Q^i_\beta$ is a certain polynomial in $x_\alpha$ of degree $\beta + \alpha_i$.

The homomorphism $\bar{\rho}_\chi$ defines a structure of $\mathfrak{g}$–module on the space of algebraic functions on the big cell, $\mathbb{C}[x_\alpha]_{\alpha \in \Delta_+}$. This module is isomorphic to the module $M^*_\chi$, which is contragradient to the Verma module over $\mathfrak{g}$ of highest weight $\chi$.

We also have another map from the Lie algebra $\mathfrak{n}_+$ to the Lie algebra of vector fields on $U \simeq N_+$, which comes from the *left* infinitesimal action of $\mathfrak{n}_+$ on its Lie group. The $i$th generator of $\mathfrak{n}_+$ maps to the vector field

$$G_i = -\frac{\partial}{\partial x_{\alpha_i}} + \sum_{\beta \in \Delta_+} R^i_\beta \frac{\partial}{\partial x_\beta},$$

where $R^i_\beta$ is a certain polynomial in $x_\alpha$ of degree $\beta - \alpha_i$. These operators generate another action of $\mathfrak{n}_+$, which commutes with the action of $\mathfrak{n}_+$, defined by $\bar{\rho}_\chi$.

Now we can construct the homomorphism $\rho$. Introduce the notation

$$A(z) = \sum_{n \in \mathbb{Z}} A(n) z^{-n-1}, \quad A \in \mathfrak{g}$$

and

$$a_\alpha(z) = \sum_{n \in \mathbb{Z}} a_\alpha(n) z^{-n-1}, \quad a^*_\alpha(z) = \sum_{n \in \mathbb{Z}} a^*_\alpha(n) z^{-n}, \quad h_i(z) = \sum_{n \in \mathbb{Z}} h_i(n) z^{-n-1}.$$

We put

$$\rho[E_i(z)] = a_{\alpha_i}(z) + \sum_{\beta \in \Delta_+} : P^i_\beta(z) a_\beta(z) :,$$

$$\rho[H_i(z)] = -\sum_{\beta \in \Delta_+} \beta(H_i) : a^*_\beta(z) a_\beta(z) : + h_i(z),$$

$$\rho[F_i(z)] = \sum_{\beta \in \Delta_+} : Q^i_\beta(z) a_\beta(z) : + h_i(z) a^*_{\alpha_i}(z) + c_i \partial_z a^*_{\alpha_i}(z),$$

where $c_i$ is a constant. Here $P_\beta^i(z)$ and $Q_\beta^i(z)$ stand for the expressions, which are obtained from the polynomials $P_\beta^i$ and $Q_\beta^i$ by insering $a_\alpha^*(z)$ instead of $x_\alpha$, and dots denote normal ordering.

We also define operators $G_i(n), i = 1, \ldots, l, n \in \mathbb{Z}$, by the generating functions

$$(4.2) \qquad G_i(z) = \sum_{n \in \mathbb{Z}} G_i(n) z^{-n-1} = -a_{\alpha_i}(z) + \sum_{\beta \in \Delta_+} : R_\beta^i(z) a_\beta(z) : .$$

**Theorem 1.** *There exist such $c_i$ in the formula for $\rho[F_i(z)], i = 1, \ldots, l$, that the map $\rho$ defines a homomorphism from the Lie algebra $\widehat{\mathfrak{g}}$ to the local completion of $U_1(\Gamma(\mathfrak{g})) \otimes U(\widetilde{\mathfrak{h}})$, which sends the central element $K \in \widehat{\mathfrak{g}}$ to $-h^\vee$.*

This was proved in [17] for $\widehat{\mathfrak{g}} = \widehat{\mathfrak{sl}}_2$ and in [18, 40] in general.

*Example.* We will write down the precise formulas in the case $\mathfrak{g} = \mathfrak{sl}_2$ [17]. We will omit the unnecessary subscripts. We have:

$$\rho[E(z)] = a(z),$$

$$\rho[H(z)] = -2 : a(z) a^*(z) : + h(z),$$

$$\rho[F(z)] = - : a(z) a^*(z) a^*(z) : + h(z) a^*(z) - 2 \partial_z a^*(z),$$

and

$$G(z) = -a(z).$$

For explicit formulas in the case $\mathfrak{g} = \mathfrak{sl}_n$, cf. [18]. □

For any $\chi(z) dz \in \mathfrak{h}^* \otimes \mathbb{C}((z)) dz$, we can consider representation $M \otimes \sigma_{\chi(z)}$ of $\Gamma(\mathfrak{g})) \oplus \widetilde{\mathfrak{h}}$. According to Theorem 1, the homomorphism $\rho$ defines on $M \otimes \sigma_{\chi(z)}$ a structure of $\widehat{\mathfrak{g}}$–module of the critical level, which we denote by $W_{\chi(z)}$. We call this module *Wakimoto module with highest weight $\chi(z)$*.

*Remark* 3. The modules, which were originally constructed in [17, 18], only depended on $\chi \in \mathfrak{h}^* \simeq \mathfrak{h}^* dz/z$. However, the generalization of that construction to arbitrary $\chi(z)$ is straightforward. □

*Remark* 4. The Laurent power series $\chi(z)$ should be considered as an element of the space of $-h^\vee$–connections on a principal $H$–bundle on the (formal) punctured disc, where $H$ is a Lie group of $\mathfrak{h}$. Locally such a connection can be written as a first order differential operator $-h^\vee \partial_z + \chi(z) dz$, where $\chi(z)$ is an element of $\mathfrak{h} \otimes \mathbb{C}((z))$, which is isomorphic to $\mathfrak{h}^* \otimes \mathbb{C}((z)) dz$ via the non-degenerate scalar product $(\cdot, \cdot)$ on $\mathfrak{h}^*$. We can view such a connection as a linear functional on $\widehat{\mathfrak{h}}$, which is equal to $-h^\vee$ on the central element. Here $\widehat{\mathfrak{h}}$ is the Heisenberg Lie algebra, which is the central extension of the Lie algebra of gauge transformations of this bundle.

The space of connections is a torsor over the space of one-forms $\mathfrak{h}^* \otimes \mathbb{C}((z)) dz$. If we trivialize the bundle, we can identify this torsor with the space of one-forms. □

The following result, which is proved in the Appendix, will be important in the next section.

**Lemma 2.** *Let $\mu(z)$ be a highest weight of the form*

$$\mu(z) = -\frac{\alpha_i}{z} + \sum_{n=0}^{\infty} \mu^{(n)} z^n, \qquad \mu^{(n)} \in \mathfrak{h}^*.$$

*The vector $G_i(-1)v \in W_{\mu(z)}$ is a singular vector of imaginary weight, if and only if $(\alpha_i, \mu^{(0)}) = 0$.*

We can also define another representation of $\widehat{\mathfrak{g}}$ at the critical level. Let $\pi_0 \simeq \mathbb{C}[h_i(n)], i = 1, \ldots, l, n > 0$, be the $\widetilde{\mathfrak{h}}$–module, on which $h_i(n), n \geq 0$, act by 0 and $h_i(n), n < 0$ act as on the free module with one generator. Then $M \otimes \pi_0$ is a $\widehat{\mathfrak{g}}$–module via the homomorphism $\rho$. We denote this module by $\mathbb{W}_0$.

One can check directly that the generating vector $\widetilde{v} = v \otimes 1$ of $\mathbb{W}_0$ is annihilated by $A(m), A \in \mathfrak{g}, m \geq 0$. Therefore the map $\rho$ defines a homomorphism $\widetilde{\rho} : \mathbb{V}_0 \to \mathbb{W}_0$, which sends $v_0 \in \mathbb{V}_0$ to $\widetilde{v} \in \mathbb{W}_0$. One can show that $\widetilde{\rho}$ is a embedding. Under this homomorphism all singular vectors of imaginary weight from $\mathbb{V}_0$ get mapped into the subspace $\pi_0 \subset \mathbb{W}_0$ [10]. Thus, we obtain an injective map $\widetilde{\rho}_{\mathcal{Z}} : \mathcal{Z}(\widehat{\mathfrak{g}}) \to \pi_0$.

In [10, 11] the image of this map was described. Let us recall this description.

We can consider the algebra $\pi_0$ as the algebra of differential polynomials. The action of derivative $\partial$ on $\pi_0$ is defined by its action on the generators: $\partial h_i(n) = -nh_i(n-1)$, and by the Leibnitz rule. The action of $\partial$ on $\mathcal{Z}(\widehat{\mathfrak{g}})$ and on $\pi_0$ commutes with the map $\widetilde{\rho}_{\mathcal{Z}}$. Therefore the image of $\mathcal{Z}(\widehat{\mathfrak{g}})$ is a differential subalgebra of $\pi_0$.

On the other hand, for any $\mathfrak{g}$ one can define a remarkable differential subalgebra of $\pi_0 = \pi_0(\mathfrak{g})$, which consists of "densities" of the corresponding classical $\mathcal{W}$–algebra, defined by the Drinfeld-Sokolov reduction [42, 43]. This subalgebra can be characterized as the space of invariants of the nilpotent subalgebra $\mathfrak{n}_+$ of $\mathfrak{g}$, which acts on $\pi_0$, cf. [44], Sect. 2.4. It can be considered as a classical limit of the vertex operator algebra of the quantum $\mathcal{W}$–algebra. We denote this subalgebra by $\mathcal{W}(\mathfrak{g})$.

Now let $\mathfrak{g}^\vee$ be the Langlands dual Lie algebra of $\mathfrak{g}$. Recall that the Cartan matrix of $\mathfrak{g}^\vee$ is the transpose of the Cartan matrix of $\mathfrak{g}$. Therefore there is an isomorphism between the Cartan subalgebras of $\mathfrak{g}$ and $\mathfrak{g}^\vee$, which preserves the scalar products. This induces an isomorphism between $\pi_0(\mathfrak{g})$ and $\pi_0(\mathfrak{g}^\vee)$. Therefore we can consider $\mathcal{W}(\mathfrak{g}^\vee)$ as a differential subalgebra of $\pi_0(\mathfrak{g})$.

**Theorem 2.** [10, 11] *The image of $\mathcal{Z}(\widehat{\mathfrak{g}})$ in $\pi_0(\mathfrak{g})$ under the homomorphism $\widetilde{\rho}_{\mathcal{Z}}$ coincides with $\mathcal{W}(\mathfrak{g}^\vee)$.*

Proposition 3 follows from this Theorem, because it is known that $\mathcal{W}(\mathfrak{g}^\vee)$ is a free commutative subalgebra of $\pi_0$ generated by $\partial^n P_i, i = 1, \ldots, l, n \geq 0$, where $\deg P_i = d_i + 1$ [43, 44]. The generator $P_i$ is the image of $S_i \in \mathcal{Z}(\widehat{\mathfrak{g}})$ from Proposition 3.

*Example.* If $\mathfrak{g} = \mathfrak{sl}_2$, then $\pi_0 = \mathbb{C}[h(n)]_{n<0}$. The image of $\mathcal{Z}(\widehat{\mathfrak{g}})$ in $\pi_0$ consists of polynomials in $\partial^m P, m \geq 0$, where

$$P = \frac{1}{4}h(-1)^2 - \frac{1}{2}h(-2).$$

This space coincides with $\mathcal{W}(\mathfrak{sl}_2)$, cf. [44], Sect. 2.1. □

The center $\mathfrak{z}(\widehat{\mathfrak{g}})$ of the local completion $U_{-h^\vee}(\widehat{\mathfrak{g}})_{\text{loc}}$ of the universal enveloping algebra of $\widehat{\mathfrak{g}}$ at the critical level was described in [10]. It consists of "local expressions" in $S_1, \ldots, S_l$, i.e. Fourier components of all differential polynomials in $S_1(z), \ldots, S_l(z)$, where $S_i(z)$ is the local current associated to $S_i$ in the vertex algebra $\mathbb{V}_0$. Therefore $\mathfrak{z}(\widehat{\mathfrak{g}})$ is isomorphic to the classical $\mathcal{W}$-algebra, associated to $\widehat{\mathfrak{g}}^\vee$ by means of the Drinfeld-Sokolov reduction, cf. [10]. This $\mathcal{W}$-algebra is the space of local functionals on a certain hamiltonian space $H(\mathfrak{g}^\vee)$ [43].

Now consider the space $\widehat{\mathcal{F}}_0$ of all Fourier components of differential polynomials in $h_1(z), \ldots, h_l(z)$, cf. [44]. It is isomorphic to the space of local functionals on the space $T(\mathfrak{g})$ of $-h^\vee$-connections $-h^\vee \partial_z + \chi(z)dz$. The map $\widetilde{\rho}_\mathcal{Z} : \mathcal{Z}(\widehat{\mathfrak{g}}) \to \pi_0$ gives us an embedding $\widetilde{\rho}_\mathfrak{z} : \mathfrak{z}(\widehat{\mathfrak{g}}) \to \widehat{\mathcal{F}}_0$. The corresponding map of "spectra" $T(\mathfrak{g}) \to H(\mathfrak{g}^\vee)$ is nothing but the generalized Miura transformation [43].

Elements of $\mathfrak{z}(\widehat{\mathfrak{g}})$ act on Wakimoto modules by multiplication by constants. One can describe the map $\widetilde{\rho}_\mathfrak{z}$ in terms of Wakimoto modules as follows: for $X \in \mathfrak{z}(\widehat{\mathfrak{g}})$, the value of $\widetilde{\rho}_\mathfrak{z}(X)$ at $-h^\vee \partial_z + \chi(z)dz$ is equal to the action of $X$ on the Wakimoto module $W_{\chi(z)}$. This shows that the map $\widetilde{\rho}_\mathfrak{z}$ is an affine analogue of the Harish-Chandra homomorphism.

Recall that the Harish-Chandra homomorphism is a map from the center $\mathfrak{z}(\mathfrak{g})$ of the universal enveloping algebra $U(\mathfrak{g})$ of $\mathfrak{g}$ to the algebra of polynomials on the dual space $\mathfrak{h}^*$ of the Cartan subalgebra $\mathfrak{h}$ of $\mathfrak{g}$. Any element of $\mathfrak{z}(\mathfrak{g})$ acts on Verma modules by multiplication by a constant. Harish-Chandra homomorphism maps a central element $X$ to the polynomial on $\mathfrak{h}^*$, whose value at $\chi$ is equal to the action of $X$ on $M_\chi$.

*Example.* If $\mathfrak{g} = \mathfrak{sl}_2$, then $\mathfrak{z}(\widehat{\mathfrak{g}})$ consists of all local expressions in the components $S_n$ of the current

$$S(z) = \frac{1}{2}\sum_{a=1}^{d} :I_a(z)I^a(z): = \sum_{n\in\mathbb{Z}} S_n z^{-n-2}.$$

It is therefore sufficient to describe the values of $\widetilde{\rho}_\mathfrak{z}$ on $S_n$. Denote

$$s(z) = \sum_{n\in\mathbb{Z}} s_n z^{-n-2} = \widetilde{\rho}_\mathfrak{z}[S(z)].$$

Then we have according to the previous example,

$$s(z) = \frac{1}{4}h(z)^2 - \frac{1}{2}\partial_z h(z).$$

Therefore the central element $s_n$ acts on the module $W_{\chi(z)}$ by multiplication by

$$\frac{1}{4} \sum_{m \in \mathbb{Z}} \chi^{(m)} \chi^{(n-m)} + \frac{n+1}{2} \chi^{(n)},$$

where we put $\chi(z) = \sum_{m \in \mathbb{Z}} \chi^{(m)} z^{-m-1}$. □

In the next section we will use Wakimoto modules and the description of the center to construct eigenvectors of the Gaudin hamiltonians and to compute their spectrum.

## 5. Bethe vectors from Wakimoto modules.

We will construct eigenvectors using invariant functionals on tensor products of Wakimoto modules. We will restrict ourselves with modules $W_{\chi(z)}$, for which $\chi(z)$ has the form

$$\chi(z) = \frac{\chi^{(-1)}}{z} + \sum_{n=0}^{\infty} \chi^{(n)} z^n, \qquad \chi^{(n)} \in \mathfrak{h}^*.$$

In other word, we consider only the modules associated to $-h^\vee$–connections on the formal disc, which are regular or have regular singularity at the origin. We will call such highest weight $\chi(z)$ *regular*.

Slightly abusing notation we will write $\chi$ instead of $\chi^{(-1)} = \operatorname{Res}_{z=0} \chi(z)$.

Let us recall the geometric construction of the Heisenberg algebra $\Gamma := \Gamma(\mathfrak{g})$, cf., e.g., [45]. Consider the spaces $\mathcal{F}^0 = U \otimes \mathbb{C}((t))$ and $\mathcal{F}^1 = U \otimes \mathbb{C}((t))dt$ of functions and one-forms on the formal punctured disc with values in the linear space $U$ with coordinates $x_\alpha, \alpha \in \Delta_+$. There is a natural non-degenerate pairing between them:

$$(\cdot, \cdot) : \mathcal{F}^0 \times \mathcal{F}^1 \to \mathbb{C},$$

which sends $(f(t), g(t)dt)$ to $(2\pi i)^{-1} \int \langle f(t), g(t) \rangle dt$. Here $\langle \cdot, \cdot \rangle$ denotes the scalar product on the linear space $U$, with respect to which $x_\alpha$'s are orthonormal. To the scalar product $(\cdot, \cdot)$ we can associate in the standard way a central extension of the commutative Lie algebra $\mathcal{F}^0 \oplus \mathcal{F}^1$. This is our Heisenberg Lie algebra $\Gamma$. The generators $a_\alpha(m), a_\alpha^*(m)$ correspond to $x_\alpha \otimes t^m \in \mathcal{F}^0$ and $x_\alpha \otimes t^{m-1}dt \in \mathcal{F}^1$, respectively.

Let $\mathbf{x} = x_1, \ldots, x_p$ be a set of distinct points on the projective line. Consider Wakimoto modules $W_{\chi_i(z)}$ with regular highest weights $\chi_i(z), i = 1, \ldots, p$.

Let us choose a global coordinate $t$ on $\mathbb{CP}^1$ and the corresponding local coordinates $t - x_i, i = 1, \ldots, p$, at our points. Denote $\Gamma(x_i) = U \otimes \mathbb{C}((t-x_i)) \oplus U \otimes \mathbb{C}((t-x_i))dt$. Let $\Gamma_p$ be the central extension of the Lie algebra $\oplus_{i=1}^p \Gamma(x_i)$ by a one-dimensional center, which coincides with our standard central extension on each of the summands. The Lie algebra $\Gamma_p$ acts on $M^{\otimes p}$ in a natural way, in particular, the central element **1** acts as the identity.

Consider the commutative Lie algebra $\mathcal{F}_\mathbf{x} := \mathcal{F}_\mathbf{x}^0 \oplus \mathcal{F}_\mathbf{x}^1$, where $\mathcal{F}_\mathbf{x}^0$ ($\mathcal{F}_\mathbf{x}^1$) is the space of $U$–valued regular functions (one-forms) on $\mathbb{CP}^1 \setminus \{x_1, \ldots, x_p\}$, which vanish (have regular singularity) at infinity. We have an embedding of $\mathcal{F}_{\mathbf{x}}$ into $\oplus_{i=1}^{p}\Gamma(x_i)$, obtained by expanding a function and a one-form at a given point. The restriction of the central extension to the image of this embedding is trivial. Therefore we can lift it to an embedding $\mathcal{F}_{\mathbf{x}} \to \Gamma_p$.

We also introduce the commutative Lie algebra $\mathfrak{h}_{\mathbf{x}}$ of $\mathfrak{h}$–valued regular functions on $\mathbb{CP}^1 \backslash \{x_1, \ldots, x_p\}$, which vanish at infinity. We have an embedding $\mathfrak{h}_{\mathbf{x}} \to \oplus_{i=1}^{p} \mathfrak{h} \otimes \mathbb{C}((t - x_i))$. Denote $\boldsymbol{\chi}(z) = \chi_1(z), \ldots, \chi_p(z)$, where $\xi(z) \in \mathfrak{h}^* \otimes \mathbb{C}((z))dz$ and let $\sigma_{\boldsymbol{\chi}(z)}$ be the one-dimensional representation of $\oplus_{i=1}^{p} \mathfrak{h} \otimes \mathbb{C}((t - x_i))$, on which $\mathfrak{h} \otimes \mathbb{C}((t - x_i))$ acts according to its character $\chi_i(z)$. Thus, $\mathfrak{h}_{\mathbf{x}}$ also acts on $\sigma_{\boldsymbol{\chi}(z)}$.

The Lie algebra $\mathcal{H}_{\mathbf{x}} := \mathcal{F}_{\mathbf{x}} \oplus \mathfrak{h}_{\mathbf{x}}$ acts on

$$M^{\otimes p} \otimes \sigma_{\boldsymbol{\chi}(z)} = \otimes_{i=1}^{p} W_{\chi_i(z)}.$$

Denote by $J_{\boldsymbol{\chi}(z)}$ the space of $\mathcal{H}_{\mathbf{x}}$–invariant linear functionals on this tensor product.

**Proposition 4.** *Suppose that*

$$\chi_i(z) = \frac{\chi_i}{z} + \sum_{n=0}^{\infty} \chi_i^{(n)} z^n, \qquad i = 1, \ldots, p.$$

*If $\chi_i(t - x_i)$ is the expansion of*

$$\chi(t) = \sum_{i=1}^{p} \frac{\chi_i}{t - x_i}$$

*at the point $x_i$, for all $i = 1, \ldots, p$, then the space $J_{\boldsymbol{\chi}(z)}$ is one-dimensional. It is generated by a functional, whose value on the tensor product of the vacuum vectors of $W_{\chi_i(z)}$ is equal to $1$.*

*Otherwise $J_{\boldsymbol{\chi}(z)} = 0$.*

*Proof.* The space $J_{\boldsymbol{\chi}(z)}$ is isomorphic to the tensor product of the dual space of the space of coinvariants of $M^{\otimes p}$ with respect to the action of $\mathcal{F}_{\mathbf{x}}$, and the space of $\mathfrak{h}_{\mathbf{x}}$–invariants of $\sigma_{\boldsymbol{\chi}(z)}$.

Since $M$ is free over the Lie subalgebra of $\Gamma$, generated by $a_\alpha(m), \alpha \in \Delta_+, m < 0$, and $a_\alpha^*(m), \alpha \in \Delta_+, m \leq 0$, we can use the same argument as in the proof of Lemma 1. We conclude that the space of coinvariants of $M^{\otimes p}$ with respect to the action of $\mathcal{F}_{\mathbf{x}}$ is one-dimensional and that the projection of the tensor product of the vacuum vectors of $W_{\chi_i(z)}$ on this space is non-trivial. Therefore there exists an $\mathcal{H}_{\mathbf{x}}$–invariant functional on $\otimes_{i=1}^{p} W_{\chi_i(z)}$, whose value on the tensor product of the vacuum vectors is equal to $1$.

Let $f(t)$ be an element of $\mathfrak{h}_{\mathbf{x}}$. This is a meromorphic $\mathfrak{h}$–valued function on $\mathbb{CP}^1$, which may have singularities only at $x_i$'s and which vanishes at infinity. It acts on the one-dimensional module $\sigma_{\boldsymbol{\chi}(z)}$ by multiplication by

$$\sum_{i=1}^{p} \frac{1}{2\pi i} \int_{J_i} \chi_i(t-z_i)[f(t)]dt,$$

where $\int_i$ stands for the integral over a small contour around $x_i$. This sum vanishes for any $f(t)$, if and only if $\chi_i(t-x_i)$ is the expansion of

$$\chi(t) = \sum_{i=1}^{p} \frac{\chi_i}{t-x_i}$$

at the point $x_i$ for all $i=1,\ldots,p$. Therefore the space of $\mathfrak{h}_{\mathbf{x}}$–invariants of $\sigma_{\boldsymbol{\chi}(z)}$ is non-zero, if and only if the condition of Proposition is satisfied. $\square$

Now fix highest weights of $\mathfrak{g}$, $\lambda_1,\ldots,\lambda_N$, and a set of simple roots of $\mathfrak{g}$, $\alpha_{i_1},\ldots,\alpha_{i_m}$. Consider the function

(5.1) $$\lambda(t) = \sum_{i=1}^{N} \frac{\lambda_i}{t-z_i} - \sum_{j=1}^{m} \frac{\alpha_{i_j}}{t-w_j}.$$

Denote by $\lambda_i(t-z_i)$ the expansions of $\lambda(t)$ at the points $z_i, i=1,\ldots,N$, and by $\mu_j(t-w_j)$ the expansions of $\lambda(t)$ at the points $w_j, j=1,\ldots,m$. We have:

$$\lambda_i(z) = \frac{\lambda_i}{z} + \ldots, \qquad \mu_j(z) = -\frac{\alpha_{i_j}}{z} + \mu_j^{(0)} + \ldots,$$

where

$$\mu_j^{(0)} = \sum_{i=1}^{N} \frac{\lambda_i}{w_j - z_i} - \sum_{s \neq j} \frac{\alpha_{i_s}}{w_j - w_s}.$$

We put $p = N+m$, and denote $x_i = z_i, \chi_i(z) = \lambda_i(z), i=1,\ldots,N; x_{N+j} = w_j, \chi_{N+j}(z) = -\mu_{i_j}(z), j=1,\ldots,m$. According to Proposition 4, the space $J_{\boldsymbol{\lambda}(z),\boldsymbol{\mu}(z)}$ of $\mathcal{H}_{\mathbf{z},\mathbf{w}}$–invariant functionals on the tensor product $\otimes_{i=1}^{N} W_{\lambda_i(z)} \otimes_{j=1}^{m} W_{\mu_j(z)}$ is one-dimensional. It is generated by a functional $\tau_{N,m}$, whose value on the tensor product of the vacuum vectors is equal to 1. This is a linear map

(5.2) $$\tau_{N,m} : \otimes_{i=1}^{N} W_{\lambda_i(z)} \otimes_{j=1}^{m} W_{\mu_j(z)} \to \mathbb{C}.$$

We will obtain eigenvectors of the operators $Z(u), Z \in \mathcal{Z}(\widehat{\mathfrak{g}})$, by restricting the functional $\tau_{N,m}$ to a certain subspace.

According to Lemma 2, the vectors $G_{i_j}(-1)v \in W_{\mu_j(z)}, j=1,\ldots,m$, are singular of imaginary weight, if and only if the following system of equations is satisfied

(5.3) $$\sum_{i=1}^{N} \frac{(\lambda_i, \alpha_{i_j})}{w_j - z_i} - \sum_{s \neq j} \frac{(\alpha_{i_s}, \alpha_{i_j})}{w_j - w_s} = 0, \quad j=1,\ldots,m.$$

These are precisely the Bethe ansatz equations.

GAUDIN MODEL, BETHE ANSATZ AND CRITICAL LEVEL 21

Denote by $\widetilde{W}_{\lambda(z)}$ the subspace of a Wakimoto module $W_{\lambda(z)}$, which is generated from the vector $v$ by the operators $a_\alpha^*(0), \alpha \in \Delta_+$. This space is stable under the action of the constant subalgebra $\mathfrak{g}$ of $\widehat{\mathfrak{g}}$, and is isomorphic to the module $M_\lambda^*$ contragradient to the Verma module over $\mathfrak{g}$ with highest weight $\lambda = \mathrm{Res}_{z=0}\, \lambda(z)$. Indeed, only the 0th components $a_\alpha(0)$ and $a_\alpha^*(0)$ will contribute to the action of $\mathfrak{g}$ on $\widetilde{W}_{\lambda(z)}$. Our formulas show that if we restrict the homomorphism $\rho$ to $\mathfrak{g}$ and replace $a_\alpha^*(0)$ by $x_\alpha$ and $a_\alpha(0)$ by $\partial/\partial x_\alpha$, we will obtain the formulas for the homomorphism $\bar\rho_\lambda$, which defines on $\mathbb{C}[x_\alpha]_{\alpha \in \Delta_+} \simeq \widetilde{W}_{\lambda(z)}$ a structure of $\mathfrak{g}$–module isomorphic to $M_\lambda^*$.

The restriction of the homomorphism $\tau_{N,m}$ (5.2) to the subspace

$$(5.4) \qquad \otimes_{i=1}^N \widetilde{W}_{\lambda_i(z)} \otimes G_{i_1}(-1)v \otimes \ldots \otimes G_{i_m}(-1)v$$

defines a linear functional

$$\psi(w_1^{i_1}, \ldots, w_m^{i_m}) : \otimes_{i=1}^N M_{\lambda_i}^* \to \mathbb{C}.$$

The $\mathfrak{g}$–module $\otimes_{i=1} M_{\lambda_i}^*$ is graded by the weights of the Cartan subalgebra $\mathfrak{h}$ of $\mathfrak{g}$, so that all homogeneous components are finite-dimensional. We will show in the proof of Lemma 3 that the functional $\psi(w_1^{i_1}, \ldots, w_m^{i_m})$ vanishes on all homogeneous components except the one of weight $\sum_{i=1}^N \lambda_i - \sum_{j=1}^m \alpha_{i_j}$. Therefore it corresponds to a vector $\phi(w_1^{i_1}, \ldots, w_m^{i_m})$ in the tensor product of the Verma modules over $\mathfrak{g}$, $\otimes_{i=1}^N M_{\lambda_i}$, of weight $\sum_{i=1}^N \lambda_i - \sum_{j=1}^m \alpha_{i_j}$.

There is a differential algebra homomorphism $r_{N,m}$ from $\pi_0$ to the algebra of rational functions in $u$, having singularities at $z_1, \ldots, z_N, w_1, \ldots, w_m$, on which the derivative acts as $\partial_u$. In order to define it we have to specify the values of this homomorphism on the elements $h_s(-1)$ of $\pi_0$. We put

$$(5.5) \qquad r_{N,m}(h_s(-1)) = h_s(u) := \sum_{i=1}^N \frac{\lambda_i(H_s)}{u - z_i} - \sum_{j=1}^m \frac{\alpha_{i_j}(H_s)}{u - w_j}.$$

This implies that

$$r_{N,m}(h_{s_1}(n_1) \ldots h_{s_M}(n_M)) = \prod_{k=1}^M \frac{1}{(-n_k - 1)!} \frac{\partial^{-n_k - 1}}{\partial u^{-n_k - 1}} h_{s_k}(u).$$

**Theorem 3.** *If Bethe ansatz equations (5.3) are satisfied, then vector $\phi(w_1^{i_1}, \ldots, w_m^{i_m})$ is an eigenvector of the operator $Z(u)$ for any $Z \in \mathcal{Z}(\widehat{\mathfrak{g}})$ with the eigenvalue $r_{N,m}\widetilde{\rho}_\mathcal{Z}(Z)$.*

In order to prove the Theorem, we will need some properties of the vertex algebra $\mathbb{W}_0$, cf. [10, 11]. Each vector $A \in \mathbb{W}_0$ defines a formal power series $A(z)$, whose coefficients are linear operators acting on representations of $\Gamma(\mathfrak{g}) \oplus \widetilde{\mathfrak{h}}$. We call these formal power series *local currents* (note that they were called *fields* in [10, 11]). They satisfy all axioms of vertex operator algebras [46, 47], except for the existence of the Virasoro element; that is why we call $\mathbb{W}_0$ a vertex algebra.

The local current, corresponding to a monomial basis element $A_1(n_1)\ldots A_m(n_m)v \in \mathbb{W}_0$, where $A_i$ stands for $a_\alpha$, $a_\alpha^*$, or $h_i$, is given by

$$\frac{1}{(-n_1 - d(A_1))!} \cdots \frac{1}{(-n_m - d(A_m))!} : \partial_z^{-n_1-d(A_1)} A_1(z) \ldots \partial_z^{-n_m-d(A_m)} A_m(z) : .$$

Here we put $d(a_\alpha) = d(h_i) = 1, d(a_\alpha^*) = 0$. This current does not depend on the order in which we apply normal ordering. In general one should use normal ordering nested from right to left.

The space $\mathbb{V}_0$ is also a vertex algebra (but not a vertex operator algebra) [48], and the homomorphism $\rho$ defines an embedding of this vertex algebra into $\mathbb{W}_0$.

Let us assign representations $L_i$ of $\Gamma(\mathfrak{g}) \oplus \widetilde{\mathfrak{h}}$ to the points $x_i, i = 1, \ldots, p$, of $\mathbb{CP}^1$. Denote by $J(L_1, \ldots, L_p)$ the space of $\mathcal{H}_{\mathbf{x}}$-invariant functionals on $\otimes_{i=1}^p L_i$.

Introduce a $\mathbb{Z}$-grading $d$ on $\mathbb{W}_0$ by putting $d(a_\alpha(n)) = d(a_\alpha^*(n)) = d(h_i(n)) = -n$. Let $Y(z)$ be the local current corresponding to an element $Y \in \mathbb{W}_0$ of degree $d(Y)$. We will identify the $(1-d(Y))$-differential $(t-x_j)^{n-d(Y)+1} dt^{1-d(Y)} \in \mathbb{C}((t-x_j))dt^{1-d(Y)}$ with the operator $Y(n)^{(j)}$ acting on the module $\otimes_{i=1}^p L_i$. Consider the space $\mathcal{F}_{\mathbf{x}}^{1-d(Y)}$ of regular $(1-d(Y))$-differentials on $\mathbb{CP}^1\backslash\{x_1, \ldots, x_p\}$, which have zero of order at least $2d(Y)-1$ at infinity. Any element $\mathcal{Y}(t)$ of this space can be expanded in powers of $t-x_i$ at each point $x_i$. The corresponding formal power series $\mathcal{Y}^{(i)}(t-x_i)$ can then be viewed as an infinite sum of operators $\sum_{m>-d(Y)} c_m Y(m)^{(i)}$ on the space $\otimes_{i=1}^p L_i$. Denote

$$\mathcal{Y} = \sum_{i=1}^p \mathcal{Y}^{(i)}(t - x_i).$$

**Proposition 5.** *For any $\tau \in J(L_1, \ldots, L_p), Y \in \mathbb{W}_0$, and $X \in \otimes_{i=1}^p L_i$, $\tau(\mathcal{Y} \cdot X) = 0$.*

In the case when $Y(z)$ is one of the elementary currents: $a_\alpha(z), a_\alpha^*(z)$, or $h_i(z)$, this follows from the definition of the space of invariant functionals. The general result follows by induction on the power of $Y$. By linearity, it is sufficient to prove the statement in the case when $\mathcal{Y}(t) = (t-x_j)^{n+d(Y)-1} dt^{1-d(Y)}, n \leq -d(Y)$. Proposition 5 is equivalent in this case to the statement that

$$\tau[Y(n)^{(j)} \cdot X] = \frac{1}{(-n-d(Y))!} \frac{\partial^{-n-d(Y)}}{\partial x_j^{-n-d(Y)}} \cdot$$
(5.6)
$$\tau\left[\sum_{i \neq j} \sum_{m=-d(Y)+1}^\infty \frac{Y(m)^{(i)}}{(x_j - x_i)^{m+d(Y)}} \cdot X\right].$$

This is a generalization of the Ward identity (3.5). Note that this identity and Proposition 5 hold for arbitrary vertex algebras.

Since each of the modules $L_i$ is automatically a $\widehat{\mathfrak{g}}$-module, the Lie algebra $\mathfrak{g}_{\mathbf{x}}$, which was introduced in Sect. 3, acts on the tensor product $\otimes_{i=1}^p L_i$.

**Corollary 1.** *For any $\tau \in J(L_1, \ldots, L_p), g \in \mathfrak{g}_\mathbf{x}$ and $X \in \otimes_{i=1}^p L_i$, $\tau(g \cdot X) = 0$.*

*Proof.* We have to apply Proposition 5.6 in the case when $Y(z) = \rho[A(z)], A \in \mathfrak{g}$. □

Corollary 1 means that any $\mathcal{H}_\mathbf{x}$-invariant functional is $\mathfrak{g}_\mathbf{x}$-invariant. This will allow us to prove Theorem 3.

*Proof of Theorem 3.* Let $\lambda_i(z)$ and $\mu_j(t)$ be the expansions of (5.1). We assign the modules $W_{\lambda_i(z)}$ to the points $z_i, i = 1, \ldots, N$, the modules $W_{\mu_j(z)}$ to the points $w_j, j = 1, \ldots, m$, and the module $\mathbb{W}_0$ to the point $u$ of $\mathbb{CP}^1$.

The module $\mathbb{W}_0$ is free over the Lie subalgebra of $\Gamma \oplus \widetilde{\mathfrak{h}}$, generated by $a_\alpha(m), \alpha \in \Delta_+, m < 0; a_\alpha(m), \alpha \in \Delta_+, m \leq 0; h_i(m), i = 1, \ldots, l, m < 0$. In the same way as in the proof of Lemma 1, we can show that the space of $\mathcal{H}_{\mathbf{z},\mathbf{w},u}$-invariant functionals is one-dimensional and is generated by a functional $\tau$, whose value on the tensor product of the vacuum vectors is equal to 1. It clearly has the property: $\tau(X, \widetilde{v}) = \tau_{N,m}(X)$ for any $X \in \otimes_{i=1}^N W_{\lambda_i(z)} \otimes_{j=1}^m W_{\mu_j(z)}$, where $\widetilde{v}$ is the generating vector of $\mathbb{W}_0$ and $\tau_{N,m}$ is the generator of the space $J_{\boldsymbol{\lambda}(z), \boldsymbol{\mu}(z)}$, cf. (5.2).

Recall that we have an embedding $\widetilde{\rho} : \mathbb{V}_0 \to \mathbb{W}_0$ of $\widehat{\mathfrak{g}}$-modules. Now let $Z \in \mathcal{Z}(\widehat{\mathfrak{g}}) \subset \mathbb{V}_0$ and consider $\tau_{N,m}(\omega, G_{i_1}(-1)v, \ldots, G_{i_m}(-1)v, \widetilde{\rho}(Z))$, where $\omega \in \otimes_{i=1}^N \widetilde{W}_{\lambda_i(z)} \subset \otimes_{i=1}^N W_{\lambda_i(z)}$. This is a linear functional of $\omega$.

We can express this linear functional in two different ways. On the one hand, we can represent $Z$ as an element of $U(\widehat{\mathfrak{g}}_-)$, i.e. in terms of $\rho[A(m)], A \in \mathfrak{g}, m < 0$. By Corollary 1 the functional $\tau$ is invariant with respect to the Lie algebra $\mathfrak{g}_{\mathbf{z},\mathbf{w},u}$. So, we can use formula (3.5) to "swap" $Z$. Since $G_{i_j}(-1)v \in W_{\mu_j(z)}, j = 1, \ldots, m$, are singular vectors of imaginary weight, $A(m) \cdot G_{i_j}(-1)v = 0$ for any $m \geq 0$. Therefore the action of the corresponding elements of the Lie algebra $\mathfrak{g}_{\mathbf{z},\mathbf{w},u}$ on these vectors is equal to 0. So, we obtain, in the same way as in the proof of Proposition 2,

$$(5.7) \quad \tau(\omega, G_{i_1}(-1)v, \ldots, G_{i_m}(-1)v, \widetilde{\rho}(Z)) = [Z(u) \cdot \psi(w_1^{i_1}, \ldots, w_m^{i_m})](\omega).$$

On the other hand, we know that $\widetilde{\rho}(Z)$ lies in $\pi_0 \subset \mathbb{W}_0$. According to (5.6),

$$\tau_{N,m}(\omega, G_{i_1}(-1)v, \ldots, G_{i_m}(-1)v, h_i(n)X) =$$

$$r_{N,m}(h_i(n))\tau_{N,m}(\omega, G_{i_1}(-1)v, \ldots, G_{i_m}(-1)v, X).$$

By applying this formula several times to $\widetilde{\rho}(Z)$, we obtain

$$(5.8)\ \tau_{N,m}(\omega, G_{i_1}(-1)v, \ldots, G_{i_m}(-1)v, \widetilde{\rho}(Z)) = [r_{N,m}\widetilde{\rho}_{\mathcal{Z}}(Z)]\psi(w_1^{i_1}, \ldots, w_m^{i_m})(\omega).$$

From (5.7) and (5.8) we obtain

$$Z(u) \cdot \phi(w_1^{i_1}, \ldots, w_m^{i_m}) = [r_{N,m}\widetilde{\rho}_{\mathcal{Z}}(Z)]\phi(w_1^{i_1}, \ldots, w_m^{i_m})$$

and Theorem follows. □

*Example.* For the element $S$ given by (3.4), we have

$$\widetilde{\rho}_{\mathcal{Z}}(S) = \frac{1}{2}\sum_{i=1}^{l} h_i(-1)h_i^*(-1) - \rho(-2).$$

Here $h_1^*,\ldots,h_l^*$, is the basis of $\mathfrak{h}$, which is dual to the basis $h_1,\ldots,h_l$ with respect to the scalar product $(\cdot,\cdot)$ on $\mathfrak{h} \simeq \mathfrak{h}^*$ and $\rho \in \mathfrak{h}$ is such that $\alpha_i(\rho) = (\alpha_i,\alpha_i)/2$. Therefore the eigenvalue of the element $S(u)$ on the vector $\phi(w_1^{i_1},\ldots,w_m^{i_m})$ is equal to

$$s_{i_1,\ldots,i_m}(u) = \frac{1}{2}\sum_{i=1}^{l} h_i(u)h_i^*(u) - \partial_u\rho(u) =$$

$$= \frac{1}{2}\left\|\sum_{i=1}^{N}\frac{\lambda_i}{u-z_i} - \sum_{j=1}^{m}\frac{\alpha_{i_j}}{u-w_j}\right\|^2 + \sum_{i=1}^{N}\frac{\lambda_i(\rho)}{(u-z_i)^2} - \frac{1}{2}\sum_{j=1}^{m}\frac{(\alpha_{i_j},\alpha_{i_j})}{(u-w_j)^2}.$$

For $\mathfrak{g} = \mathfrak{sl}_2$ this formula coincides with (2.6). It was written in [20] (cf. formula (1.24)). Using this formula we can find the eigenvalues $s_i(w_1^{i_1},\ldots,w_m^{i_m})$ of the operators $\Xi_i$ given by (1.1) on the vector $\phi(w_1^{i_1},\ldots,w_m^{i_m})$. Since $\Xi_i = \operatorname{Res}_{u=z_i} S(u)$, this eigenvalue is equal to $\operatorname{Res}_{u=z_i} s_{i_1,\ldots,i_m}(u)$. Thus, we obtain:

(5.9) $$s_i(w_1^{i_1},\ldots,w_m^{i_m}) = \sum_{j\neq i}\frac{(\lambda_i,\lambda_j)}{z_i - z_j} - \sum_{s=1}^{m}\frac{(\lambda_i,\alpha_{i_s})}{z_i - w_s}. \quad \square$$

**Lemma 3.** *The vector $\phi(w_1^{i_1},\ldots,w_m^{i_m})$ coincides up to a sign with the Bethe vector $|w_1^{i_1},\ldots,w_m^{i_m}\rangle$ given by (2.8).*

*Proof.* We will find an explicit formula for

(5.10) $$\psi(w_1^{i_1},\ldots,w_m^{i_m})(\omega) = \tau_{N,m}(\omega,G_{i_1}v,\ldots,G_{i_m}v).$$

and compare it with formula (2.8) for $|w_1^{i_1},\ldots,w_m^{i_m}\rangle$.

This computation is essentially equivalent to the computation of a bosonic correlation function from [35]. For $X \in M_\chi^*$ denote by $P_X$ the corresponding polynomial in $x_\alpha$'s. Let $P_X(z)$ be the local current, which is obtained from $P_X$ by replacing $x_\alpha$ with $a_\alpha^*(z)$. In conformal field theory language,

$$\tau_{N,m}(X_1,\ldots,X_N,G_{i_1}v,\ldots,G_{i_m}v),$$

where $X_i \in M_{\lambda_i}^*$, is the bosonic $\beta\gamma$ correlation function

(5.11) $$\left\langle \prod_{i=1}^{N} P_{X_i}(z_i) \prod_{j=1}^{m} G_{i_j}(w_j) \right\rangle.$$

We will use the generalized Ward identity (5.6) in the case when $\mathcal{Y} = \frac{G_{\beta_j}}{t-w_j}$, where $G_{\beta_j}$ is a root generator of the left nilpotent subalgebra $\mathfrak{n}_+$ of $\mathfrak{g}$.

This identity reads

$$\tau_{N,m}\left(\omega, G_{\beta_1}(-1)v, \ldots, \underset{j}{G_{\beta_j}(-1)v}, \ldots, G_{\beta_m}(-1)v\right) =$$

(5.12)
$$= \sum_{i=1}^{N} \frac{1}{w_j - z_i} \tau_{N,m}\left(G_{\beta_j}^{(i)}\omega, G_{\beta_1}(-1)v, \ldots, \underset{j}{v}, \ldots, G_{\beta_m}(-1)v\right) +$$

$$+ \sum_{s=1}^{m} \frac{c_{\beta_j,\beta_s}^{\beta_j+\beta_s}}{w_j - w_s} \tau_{N,m}\left(\omega, G_{\beta_1}(-1)v, \ldots, \underset{j}{v}, \ldots, \underset{s}{G_{\beta_j+\beta_s}(-1)v}, \ldots G_{\beta_m}(-1)v\right),$$

where $\omega \in \otimes_{i=1}^{N}\widetilde{W}_{\lambda_i(z)} \simeq \otimes_{i=1}^{N} M_{\lambda_i}^*$ and $c_{\beta,\gamma}^{\alpha}$ are the structure constants in the nilpotent subalgebra $\mathfrak{n}_+$ of $\mathfrak{g}$.

By successive use of this identity, we obtain:

$$\tau_{N,m}(\omega, G_{i_1}v, \ldots, G_{i_m}v) = \tau_{N,m}(\omega', v, \ldots, v)$$

for some $\omega' \in \otimes_{i=1}^{N} M_{\lambda_i}^*$.

Recall that there is a linear pairing $\langle \cdot, \cdot \rangle : M_\chi \times M_\chi^* \to \mathbb{C}$, such that $\langle F_i Y, X \rangle = \langle Y, E_i X \rangle$, $i = 1, \ldots, l$, cf. (3.3). Denote by $\jmath : M_\chi^* \to \mathbb{C}$ the pairing with the highest weight vector $x_\chi$ of $M_\chi$: $\jmath(X) = \langle x_\chi, X \rangle$.

When $\mathcal{Y} = a_\alpha^*$, the Ward identity (5.6) reads as follows:

$$\tau_{N,m}(a_\alpha^*(0)\omega, v, \ldots, v) = 0, \qquad \alpha \in \Delta_+.$$

Therefore
$$\tau_{N,m}(\omega', v, \ldots, v) = \jmath(\omega').$$

If $\omega$ is homogeneous of weight $\gamma$, then $\omega'$ is also homogeneous of weight $\gamma + \sum_{j=1}^{m} \alpha_{i_j}$. But $\jmath(\omega') = 0$, if the weight of $\omega'$ is not equal to $\sum_{i=1}^{N} \lambda_i$. Therefore $\psi(w_1^{i_1}, \ldots, w_m^{i_m})(\omega) = 0$, if the weight of $\omega$ is not equal to $\sum_{i=1}^{N} \lambda_i - \sum_{j=1}^{m} \alpha_{i_j}$.

Following [35], we obtain by induction the following formula:

$$\psi(w_1^{i_1}, \ldots, w_m^{i_m})(X_1, \ldots, X_N) = \sum_{p=(I^1,\ldots,I^N)} \prod_{j=1}^{N} \frac{\jmath(G_{i_1^j} \ldots G_{i_{a_j}^j} P_{X_j})}{(w_{i_1^j} - w_{i_2^j}) \ldots (w_{i_{a_j}^j} - z_j)},$$

where we used (5.10) and notation from Sect. 3.

But
$$\jmath(G_{i_1} \ldots G_{i_m} P) = (-1)^m \jmath(E_{i_m} \ldots E_{i_1} P)$$

(cf. Lemma 3.3 from [35]). Therefore we have:

$$\psi(w_1^{i_1}, \ldots, w_m^{i_m})(X_1, \ldots, X_N) =$$

(5.13)
$$(-1)^m \sum_{p=(I^1,\ldots,I^N)} \prod_{j=1}^{N} \frac{\jmath(E_{i_{a_j}^j} \ldots E_{i_1^j} P_{X_j})}{(w_{i_1^j} - w_{i_2^j}) \ldots (w_{i_{a_j}^j} - z_j)}.$$

On the other hand, the pairing of a monomial basis element $F_{i_1} \ldots F_{i_j} x_\chi$ and $X \in M_\chi^*$ is equal to

$$\langle x_\lambda, E_{i_j} \ldots E_{i_1} X \rangle = \jmath(E_{i_j} \ldots E_{i_1} P_X).$$

Thus, we see that the vector $|w_1^{i_1}, \ldots, w_m^{i_m}\rangle \in \otimes_{i=1}^N M_{\lambda_i}$ given by (2.8) defines a linear functional on $\otimes_{i=1}^N M_{\lambda_i}^*$, which coincides with the functional $(-1)^m \psi(w_1^{i_1}, \ldots, w_m^{i_m})$. Therefore $\phi(w_1^{i_1}, \ldots, w_m^{i_m}) = (-1)^m |w_1^{i_1}, \ldots, w_m^{i_m}\rangle$. □

There is an interesting "analytic" interpretation of Bethe ansatz equations (5.3).

**Proposition 6.** *The Bethe ansatz equations (5.3) are satisfied if and only if the eigenvalue of any $Z \in \mathcal{Z}(\widehat{\mathfrak{g}})$ on the vector $\phi(w_1^{i_1}, \ldots, w_m^{i_m})$ is non-singular at the points $w_1, \ldots, w_m$.*

*Proof.* Denote by $\mathfrak{h}_i^\perp$ the orthogonal complement to the one-dimensional subspace of $\mathfrak{h}$ generated by $H_i \in \mathfrak{h}$. Polynomials in $h_i(n), n < 0$, form a subspace $\pi_{0,i}$ of $\pi_0$. Polynomials in $h(n), h \in \mathfrak{h}_i^\perp, n < 0$, form a subspace $\pi_{0,i}^\perp$ of $\pi_0$. We have: $\pi_0(\mathfrak{g}) = \pi_{0,i} \otimes \pi_{0,i}^\perp$.

The space $\pi_{0,i}$ is isomorphic to $\pi_0(\mathfrak{sl}_2)$. Denote by $\mathcal{W}_i \subset \pi_{0,i}$ the subspace corresponding to $\mathcal{W}(\mathfrak{sl}_2) \subset \pi_0(\mathfrak{sl}_2)$.

Now fix $j = 1, \ldots, m$. The image of $\mathcal{Z}(\widehat{\mathfrak{g}})$ in $\pi_0$ is contained in the tensor product $\pi_{0,i_j}^\perp \otimes \mathcal{W}_{i_j}$ [10]. Therefore $\widetilde{\rho}_\mathcal{Z}(Z)$ can be decomposed as $\sum_m X_m Y_m$, where $X_m \in \pi_{0,i_j}^\perp, Y_m \in \mathcal{W}_{i_j}$. Then we have:

$$[r_{N,m} \widetilde{\rho}_\mathcal{Z}](Z) = \sum_m r_{N,m}(X_m) \cdot r_{N,m}(Y_m).$$

From the definition of the homomorphism $r_{N,m}$ it is clear that $r_{N,m}(X_m)$ is non-singular at $u = w_j$. We will show now that $r_{N,m}(Y)$ is non-singular at $u = w_j$ for any $Y \in \mathcal{W}_{i_j}$ if and only if the Bethe ansatz equation (5.3) is satisfied.

We know that $\mathcal{W}_{i_j}$ is the space of differential polynomials in

$$P_{i_j} = \frac{1}{2} h_{i_j}(-1)^2 - h_{i_j}(-2).$$

Since $r_{N,m}$ is a homomorphism of differential algebras, it is sufficient to check that $r_{N,m}(P_{i_j})$ is non-singular at $u = w_j$, if and only if the Bethe ansatz equation is satisfied. Applying formula (5.5) we obtain that $r_{N,m}(P_{i_j})$ is equal to

$$\frac{1}{2} \left( \sum_{i=1}^N \frac{\lambda_i(H_{i_j})}{u - z_i} - \sum_{s=1}^m \frac{\alpha_{i_s}(H_{i_j})}{u - w_s} \right)^2 - \partial_u \left( \sum_{i=1}^N \frac{\lambda_i(H_{i_j})}{u - z_i} - \sum_{s=1}^m \frac{\alpha_{i_s}(H_{i_j})}{u - w_s} \right).$$

The possible singular terms at $u = w_j$ are

$$\frac{(\alpha_{i_j}(H_{i_j}))^2/2 - \alpha_{i_j}(H_{i_j})}{(u - w_j)^2},$$

which vanishes because $\alpha_{i_j}(H_{i_j}) = 2$, and

$$\frac{1}{u-w_j} \cdot \frac{4}{(\alpha_{i_j}, \alpha_{i_j})} \left( -\sum_{i=1}^{N} \frac{(\lambda_i, \alpha_{i_j})}{u-z_i} + \sum_{s \neq j} \frac{(\alpha_{i_s}, \alpha_{i_j})}{u-w_s} \right)$$

(note that $\lambda(H_{i_j}) = 2(\lambda, \alpha_{i_j})/(\alpha_{i_j}, \alpha_{i_j})$). The latter is non-singular at $u = w_j$, if and only if the Bethe ansatz equation (5.3) is satisfied. □

*Remark* 5. Proposition 6 can be considered as a guiding principle for finding eigenvalues of Gaudin's hamiltonians. This *analytic Bethe ansatz* has been successfully applied to various models of statistical mechanics, cf., e.g., [49]. □

In conclusion of this section, let us remark that the Gaudin model can be generalized to an arbitrary Kac-Moody Lie algebra $\mathfrak{g}$ associated to a symmetrizable Cartan matrix.

Indeed, for such $\mathfrak{g}$ we can choose the bases $\{I_a\}$ and $\{I^a\}$, which are orthogonal to each other with respect to the invariant scalar product, cf. [6]. Then the operators $\Xi_i$ given by (1.1) will be well-defined on the $N$–fold tensor product of $\mathfrak{g}$–modules from the category $\mathcal{O}$. For instance, we can take the tensor product of integrable representations $V_{\lambda_1} \otimes \ldots \otimes V_{\lambda_N}$, or Verma modules $M_{\lambda_1} \otimes \ldots \otimes M_{\lambda_N}$.

Moreover, formula (2.8) for the Bethe vector and Bethe ansatz equations make perfect sense in this general context. So does formula (5.9). This leads us to the following conjecture.

**Conjecture.** *For an arbitrary symmetrizable Kac-Moody algebra $\mathfrak{g}$, the vector $|w_1^{i_1}, \ldots, w_m^{i_m}\rangle \in M_{\lambda_1} \otimes \ldots \otimes M_{\lambda_N}$ given by (2.8) is an eigenvector of the operators $\Xi_i, i = 1, \ldots, N$, with the eigenvalues $s_i(w_1^{i_1}, \ldots, w_m^{i_m})$ given by (5.9), if the Bethe ansatz equations (5.3) are satisfied.*

This conjecture can be proved directly using methods of [31], Sect. 7.4-7.8 (note that the proof in [31] works for an arbitrary Kac-Moody algebra). One may also hope to prove this conjecture by developing theory of Wakimoto modules for general Kac-Moody algebras.

## 6. Connection with KZ equation.

We fix $k \neq -h^\vee$. Let $\omega := \omega(z_1, \ldots, z_N)$ be a function with values in $V_{(\lambda)}$. The KZ equation is the system of partial differential equations with regular singularities on $\omega$:

(6.1) $$(k + h^\vee) \frac{\partial \omega}{\partial z_i} = \Xi_i \omega, \qquad i = 1, \ldots, N,$$

where $\Xi_i$ is given by formula (1.1).

28    BORIS FEIGIN, EDWARD FRENKEL, AND NIKOLAI RESHETIKHINThis equation can be considered as the equation on the space $H^k_{(\lambda)}$, which was defined in Sect. 3 (for more details, cf. [25]).

Consider the local current $S(z)$, corresponding to the element $S$ of the vertex operator algebra $\mathbb{V}_0$ given by (3.4),

$$(6.2) \qquad S(z) = \frac{1}{2} \sum_{a=1}^{d} :I_a(z)I^a(z): = (k+h^\vee) \sum_{n\in\mathbb{Z}} L_n z^{-n-2}.$$

It is known that the operators $L_n$ generate an action of the Virasoro algebra on any representation of $\widehat{\mathfrak{g}}$ of level $k$. They have the following commutation relations with generators of $\widehat{\mathfrak{g}}$:

$$(6.3) \qquad [L_n, A(m)] = -m A(n+m), \qquad A \in \mathfrak{g}.$$

Let $C_N$ be the space $\mathbb{C}^N$ with coordinates $z_1, \ldots, z_N$ without the diagonals and $C'_N$ be the space $\mathbb{C}^{N+1}$ with coordinates $z_1, \ldots, z_N, t$ without the diagonals. Denote by $\mathcal{B}$ and $\mathcal{B}'$ the algebras of regular functions on $C_N$ and $C'_N$, respectively.

Denote by $\mathbb{V}^{k*}_{(\lambda)}$ the dual space to $\mathbb{V}^k_{(\lambda)}$.

Introduce the $\mathcal{B}$–modules $\mathbb{V}^{k*}_{(\lambda)}(\mathbf{z}) = \mathbb{V}^{k*}_{(\lambda)} \otimes_\mathbb{C} \mathcal{B}$, $\widehat{\mathfrak{g}}_N(\mathbf{z}) = \widehat{\mathfrak{g}}_N \otimes_\mathbb{C} \mathcal{B}$, and $\mathfrak{g}'_\mathbf{z} = \mathfrak{g} \otimes_\mathbb{C} \mathcal{B}'$. The Lie algebra $\widehat{\mathfrak{g}}_N(\mathbf{z})$ naturally acts on $\mathbb{V}^{k*}_{(\lambda)}(\mathbf{z})$. The Lie algebra $\mathfrak{g}'_\mathbf{z}$ embeds into $\widehat{\mathfrak{g}}_N(\mathbf{z})$ in the same way as the Lie algebra $\mathfrak{g}_\mathbf{z}$ embeds into $\widehat{\mathfrak{g}}_N$, cf. Sect. 3. Hence, $\mathfrak{g}'_\mathbf{z}$ acts on $\mathbb{V}^{k*}_{(\lambda)}(\mathbf{z})$. The space of invariants of this action, $H^k_{(\lambda)}(\mathbf{z})$, is a $\mathcal{B}$–module. By Lemma 1, we can identify $H^k_{(\lambda)}(\mathbf{z})$ with $V^*_{(\lambda)} \otimes_\mathbb{C} \mathcal{B}$.

Since $\partial/\partial z_i, i=1,\ldots,N$, are derivations of $\mathcal{B}$, they act on the $\mathcal{B}$–modules $\widehat{\mathfrak{g}}_N(\mathbf{z})$, $\mathfrak{g}'_\mathbf{z}$, and $\mathbb{V}^{k*}_{(\lambda)}(\mathbf{z})$.

Denote by $L^{(i)*}_{-1}$ the operator on $\mathbb{V}^{k*}_{(\lambda)}(\mathbf{z})$, which acts as the operator dual to $L_{-1}$ on the $i$th factor and as the identity on all other factors.

**Lemma 4.** *The operators*

$$\nabla_i = \frac{\partial}{\partial z_i} - L^{(i)*}_{-1}, \qquad i = 1, \ldots, N,$$

*acting on the space $\mathbb{V}^{k*}_{(\lambda)}(\mathbf{z})$, commute with each other and normalize the action of the Lie algebra $\mathfrak{g}'_\mathbf{z}$.*

*Proof.* We have:

$$[\nabla_i, \nabla_j] = -\frac{\partial}{\partial z_i} L^{(j)*}_{-1} + \frac{\partial}{\partial z_j} L^{(i)*}_{-1} + [L^{(i)*}_{-1}, L^{(j)*}_{-1}].$$

The first two terms in this commutator vanish, because the operators $L^{(i)*}_{-1}$ do not depend on $z_j$, and the last term vanishes, because the operators $L^{(i)*}_{-1}$ and $L^{(j)*}_{-1}$ act on different factors of $\mathbb{V}^{k*}_{(\lambda)}(\mathbf{z})$.

To prove the second statement, consider

(6.4) $$\frac{A}{(t-z_j)^n} \in \mathfrak{g}'_{\mathbf{z}}, \qquad A \in \mathfrak{g}.$$

We have to show that the commutator of this element with $\nabla_i, i = 1, \ldots, N$, is again an element of $\mathfrak{g}'_{\mathbf{z}}$. This will prove the Lemma, because such elements generate $\mathfrak{g}'_{\mathbf{z}}$.

The action of the element (6.4) on $\mathbb{V}^k_{(\boldsymbol{\lambda})}(\mathbf{z})$ is given by

$$A_n^j := A(-n)^{(j)} - \frac{1}{(n-1)!} \frac{\partial^{n-1}}{\partial z_j^{n-1}} \sum_{s \neq i} \sum_{m=0}^{\infty} \frac{A(m)^{(s)}}{(z_j - z_s)^{m+1}}$$

(cf. (3.5)). Straightforward computation using (6.3) gives:

$$\frac{\partial}{\partial z_i} A_n^j = [A_n^j, L_{-1}^{(i)}] + \delta_{i,j} \, n A_{n+1}^j.$$

Then for the dual operators, acting on $\mathbb{V}^{k*}_{(\boldsymbol{\lambda})}(\mathbf{z})$, we obtain

$$[\frac{\partial}{\partial z_i} - L_{-1}^{(i)*}, A_n^{j*}] = \delta_{i,j} \, n A_{n+1}^{j*},$$

and Lemma follows. $\square$

Lemma 4 means that the operators $\nabla_i$ define a flat connection on the trivial bundle over $C_N$ with the fiber $H^k_{(\boldsymbol{\lambda})} \simeq V^*_{(\boldsymbol{\lambda})}$. Recall that we consider on $V^*_{(\boldsymbol{\lambda})}$ the contragradient structure of $\mathfrak{g}$–module, with respect to which it is isomorphic to $V_{(\boldsymbol{\lambda})}$.

In order to find an explicit formula for this connection, consider for any $\eta \in V^*_{(\boldsymbol{\lambda})}$ the corresponding $\mathfrak{g}_{\mathbf{z}}$–invariant functional $\tilde{\eta} \in H^k_{(\boldsymbol{\lambda})}$. By definition, $[L_{-1}^{(i)*} \cdot \tilde{\eta}](\omega) = \tilde{\eta}(L_{-1}^{(i)} \cdot \omega)$ for any $\omega \in V_{(\boldsymbol{\lambda})} \subset \mathbb{V}^k_{(\boldsymbol{\lambda})}$.

Using the fact that $A(n)^{(i)} \cdot \epsilon_N^k(\omega) = 0$ for $n > 0$ we obtain

$$L_{-1}^{(i)} \cdot \omega = \frac{1}{k+h^{\vee}} \sum_{a=1}^{d} I_a(-1)^{(i)} I^a(0)^{(i)} \cdot \omega.$$

The Ward identity (3.5) gives:

$$\tilde{\eta}(L_{-1}^{(i)} \cdot \omega) = \tilde{\eta}\left(\frac{1}{k+h^{\vee}} \sum_{j \neq i} \sum_{a=1}^{d} \frac{I^a(0)^{(i)} I_a(0)^{(j)}}{z_i - z_j} \cdot \omega\right) =$$

$$\left[\frac{1}{k+h^{\vee}} \sum_{j \neq i} \sum_{a=1}^{d} \frac{\imath(I^a)^{(i)} \imath(I_a)^{(j)}}{z_i - z_j} \cdot \eta\right](\omega) = \left[\frac{\Xi_i}{k+h^{\vee}} \cdot \eta\right](\omega).$$

Thus, the action of $\nabla_i$ on the bundle of invariants is given by

$$\nabla_i = \frac{\partial}{\partial z_i} - \frac{\Xi_i}{k+h^{\vee}}.$$

Therefore the flat sections of this bundle are solutions of the KZ equation (6.1).

Note that this construction can be carried out for arbitrary representations of $\mathfrak{g}$ instead of $V_{\lambda_i}$'s. In particular, we can take the contragradient Verma modules $M^*_{\lambda_i}$, with arbitrary highest weights $\lambda_i, i = 1 \ldots, N$. Then the flat sections of the bundle of invariants will be solutions of the KZ equation with values in $\otimes_{i=1}^N M_{\lambda_i}$. A similar analysis of the KZ equation and its generalizations can be found in [26].

Schechtman and Varchenko [30, 31] have found integral solutions of the KZ equation with values in $\otimes_{i=1}^N M_{\lambda_i}$ and $\otimes_{i=1}^N M^*_{\lambda_i}$.

We will now derive the first Schechtman-Varchenko solution using Wakimoto modules of non-critical level. Essentially this has been done by Awata, Tsuchiya and Yamada in [35]. This derivation provides a clear demonstration of how Bethe vectors appear in solutions of the KZ equation.

The construction of the Wakimoto modules from the previous section can be generalized to an arbitrary level $k$ [17, 18]. Introduce the Heisenberg Lie algebra $\widehat{\mathfrak{h}}$ with generators $b_i(n), i = 1, \ldots, l, n \in \mathbb{Z}$, and the central element $\mathbf{1}$, with the commutation relations

$$[b_i(n), b_j(m)] = (k + h^\vee) n \langle H_i, H_j \rangle \delta_{n,-m} \mathbf{1}.$$

For $\chi \in \mathfrak{h}^*$ denote by $\pi_\chi$ the Fock representation of $\widehat{\mathfrak{h}}$, which is generated by the operators $b_i(n), n < 0$ from the vacuum vector $v_\chi$, which satisfies

$$b_i(n) v_\chi = 0, \quad n > 0, \qquad b_i(0) v_\chi = \chi(H_i) v_\chi$$

for all $i = 1, \ldots, l$, and $\mathbf{1} v_\chi = v_\chi$.

We define a homomorphism $\rho_k$ from $\widehat{\mathfrak{g}}$ to the local completion of $U_1(\Gamma) \otimes U(\widehat{\mathfrak{h}})$ by replacing $h_i(z)$ with $b_i(z)$ in the formulas for the homomorphism $\rho$ from Sect. 4 and shifting by $(k + h^\vee)$ the constant $c_i$ in the formula for $\rho[F_i(z)]$. Under the homomorphism $\rho_k$ the central element $K \in \widehat{\mathfrak{g}}$ maps to $k$ [17, 18]. Denote the corresponding representation of $\widehat{\mathfrak{g}}$ in $M \otimes \pi_\chi$ by $W_{\chi,k}$.

Let $C_p$ be the space $\mathbb{C}^p$ with coordinates $x_1, \ldots, x_p$ without the diagonals and $C'_p$ be the space $\mathbb{C}^{p+1}$ with coordinates $x_1, \ldots, x_p, t$ without the diagonals. Denote by $\mathcal{A}$ and $\mathcal{A}'$ the algebras of regular functions on $C_p$ and $C'_p$, respectively.

Put $W^*_{\boldsymbol{\chi},k}(\mathbf{x}) = \otimes_{i=1}^p W^*_{\chi_i,k} \otimes_{\mathbb{C}} \mathcal{A}$, $\Gamma_p(\mathbf{x}) = \Gamma_p \otimes_{\mathbb{C}} \mathcal{A}$ and $\mathcal{H}_\mathbf{x} = (U \otimes \mathcal{A}') \oplus (U \otimes \mathcal{A}' dt) \oplus (\mathfrak{h} \otimes \mathcal{A}')$. The Lie algebra $\Gamma_p(\mathbf{z})$ acts on $W^*_{\boldsymbol{\chi},k}(\mathbf{x})$; $\mathcal{H}_\mathbf{x}$ is a Lie subalgebra of $\Gamma_p(\mathbf{x})$, hence it also acts on $W^*_{\boldsymbol{\chi},k}(\mathbf{x})$. Denote by $J^k_p(\mathbf{x})$ the space of invariants of this action. This space is a free module over $\mathcal{A}$, which is generated by the $\mathcal{H}_\mathbf{x}$–invariant $\mathcal{A}$–valued functional $\vartheta_p$ on $\otimes_{i=1}^p W_{\chi_i,k}$, whose value on $\mathbf{v}_p := v_{\chi_1} \otimes \ldots \otimes v_{\chi_p}$ is equal to 1.

On the module $W^*_{\boldsymbol{\chi},k}(\mathbf{x})$, we have a natural action of the operators

$$(6.5) \qquad \frac{\partial}{\partial x_i} - L^{(i)*}_{-1}, \qquad i = 1, \ldots, p.$$

In the same way as in the proof of Lemma 4 we can show that these operators commute with each other and normalize the action of the Lie algebra $\mathcal{H}_{\mathbf{x}}$. Thus, these operators act on the space of invariants $J_p^k(\mathbf{x})$.

In order to find an explicit formula for the action of the operators (6.5) on $\vartheta_p$, we have to compute the action of the operator $L_{-1}$ on $v_\lambda \in \pi_\lambda \subset W_{\lambda,k}$.

It is known that the action of $L_{-1}$ on $\pi_\lambda \subset W_{\lambda,k}$ coincides with the action of the operator

$$\frac{1}{k+h^\vee} \sum_{n \in \mathbb{Z}} \sum_{r=1}^{l} b_r(n) b^r(-1-n),$$

where $b^r(n)$ are the dual generators to $b_r(-n)$:

$$[b^r(n), b_s(m)] = n \delta_{r,s} \delta_{n,-m} \mathbf{1}.$$

Applying this operator to $v_\lambda \in \pi_\lambda$ we obtain

$$\frac{1}{k+h^\vee} \sum_{r=1}^{l} b_r(-1) b^r(0) v_\lambda.$$

By Ward identity (5.6),

$$[L_{-1}^{(i)*} \cdot \vartheta_p](y) = \vartheta_p(L_{-1}^{(i)} \cdot y) = \vartheta_p \left( \frac{1}{k+h^\vee} \sum_{j \neq i} \sum_{r=1}^{l} \frac{b_r(0)^{(j)} b^r(0)^{(i)}}{x_i - x_j} \cdot y \right) =$$

$$= \frac{1}{k+h^\vee} \sum_{j \neq i} \frac{(\chi_i, \chi_j)}{x_i - x_j} \vartheta_p(y).$$

Consider the following system of equations:

(6.6) $$\frac{\partial f}{\partial x_i} = \frac{1}{k+h^\vee} \sum_{j \neq i} \frac{(\chi_i, \chi_j)}{x_i - x_j} f.$$

The unique up to a constant factor solution of this system is

$$f = \prod_{i<j} (x_i - x_j)^{(\chi_i, \chi_j)/(k+h^\vee)}.$$

Put $\tau_p^k(\mathbf{x}) = f \cdot \vartheta_p$. This is an $\mathcal{H}_{\mathbf{x}}$-invariant element of $W_{\chi,k}^*(\mathbf{x})$, which satisfies:

(6.7) $$\left( \frac{\partial}{\partial x_i} - L_{-1}^{(i)*} \right) \tau_p^k(\mathbf{x}) = 0.$$

Now put $p = N+m$, $x_i = z_i, \chi_i = \lambda_i, i = 1, \ldots, N$, and $x_{N+j} = w_j, \chi_{N+j} = -\alpha_{i_j}, j = 1, \ldots, m$. Then we have:

(6.8) $$\tau_{N,m}^k(\mathbf{z}, \mathbf{w})(\mathbf{v}_{N,m}) = \prod_{i<j} (z_i - z_j)^{(\lambda_i, \lambda_j)/(k+h^\vee)} \ell,$$

where
$$\ell = \prod_{i,j}(z_i - w_j)^{-(\lambda_i, \alpha_{i_j})/(k+h^\vee)} \prod_{s<j}(w_s - w_j)^{(\alpha_{i_s}, \alpha_{i_j})/(k+h^\vee)}.$$

Denote by $\widetilde{W}_\lambda$ the subspace of $W_{\lambda,k}$, which is generated from vector $v_\lambda$ by the operators $a^*_\alpha(0)$. As a module over the constant subalgebra $\mathfrak{g}$ of $\widehat{\mathfrak{g}}$, $\widetilde{W}_\lambda$ is isomorphic to $M^*_\lambda$, cf. Sect. 5.

The restriction of $\tau^k_{N,m}(\mathbf{z}, \mathbf{w})$ to the subspace
$$\otimes_{i=1}^N \widetilde{W}_{\lambda_i} \otimes G_{i_1}(-1)v_{-\alpha_{i_1}} \otimes \ldots \otimes G_{i_m}(-1)v_{-\alpha_{i_m}}$$
defines a linear functional
$$\Psi^k(\mathbf{z}, w_1^{i_1}, \ldots, w_m^{i_m}) : \otimes_{i=1}^N M^*_{\lambda_i} \to \mathbb{C}.$$
This functional vanishes on all homogeneous components except the one of weight $\sum_{i=1}^N \lambda_i - \sum_{j=1}^m \alpha_{i_j}$. Therefore it corresponds to a vector
$$\Phi^k(\mathbf{z}, w_1^{i_1}, \ldots, w_m^{i_m}) \in \otimes_{i=1}^N M_{\lambda_i}$$
of weight $\sum_{i=1}^N \lambda_i - \sum_{j=1}^m \alpha_{i_j}$, depending on $z_1, \ldots, z_N, w_1, \ldots, w_m$.

Using the Ward identity (3.5) in the same way as in the proof of Proposition 3, we obtain:
$$\Psi^k(\mathbf{z}, w_1^{i_1}, \ldots, w_m^{i_m}) = [\tau^k_{N,m}(\mathbf{z}, \mathbf{w})(\mathbf{v}_{N,m})]\psi(w_1^{i_1}, \ldots, w_m^{i_m})$$
$$= \prod_{i<j}(z_i - z_j)^{(\lambda_i, \lambda_j)/(k+h^\vee)}\ell \; \psi(w_1^{i_1}, \ldots, w_m^{i_m}),$$
by (6.8). Hence, by Lemma 3,

(6.9) $\quad \Phi^k(\mathbf{z}, w_1^{i_1}, \ldots, w_m^{i_m}) = (-1)^m \prod_{i<j}(z_i - z_j)^{(\lambda_i, \lambda_j)/(k+h^\vee)}\ell \; |w_1^{i_1}, \ldots, w_m^{i_m}\rangle.$

Denote by $C_{m,\mathbf{z}}$ the space $\mathbb{C}^m$ with coordinates $w_1, \ldots, w_m$ without all diagonals $w_j = w_i$ and all hyperplanes of the form $w_j = z_i$. The multi-valued function $\ell$ defines a one-dimensional local system $\mathcal{L}$ on the space $C_{m,\mathbf{z}}$. It is clear that for fixed $\mathbf{z}$ the value of $\tau^k_{N,m}(\mathbf{z}, \mathbf{w})$ on any vector of the tensor product $\otimes_{i=1}^N W_{\lambda_i,k} \otimes_{j=1}^m W_{-\alpha_{i_j},k}$ is a product of $\ell$ and a rational function on $C_{m,\mathbf{z}}$.

Denote by $\Omega(\mathcal{L}) = \oplus_{i\geq 0}\Omega^i(\mathcal{L})$ the twisted de Rham complex of $\mathcal{L}$. It consists of differential forms on $C_{m,\mathbf{z}}$ with differential $d'$ acting on a form $\omega$ by the formula
$$d' \cdot \omega = d \cdot \omega + d \cdot \log \ell.$$
It is therefore convenient to represent an element $\omega$ of the twisted de Rham complex by a multivalued differential form $\ell \cdot \omega$. The action of $d'$ on $\omega$ coincides with the action of $d$ on $\ell \cdot \omega$.

If $f$ is a holomorphic function on $C_{m,\mathbf{z}}$, consider the $m$–form $\ell f \, dw_1 \ldots dw_m \in \Omega^m(\mathcal{L})$. Since it is holomorphic, it defines an element of the $m$th cohomology group $H^m(C_{m,\mathbf{z}}, \mathcal{L})$ of the de Rham complex.

There is a natural pairing between this group and the $m$th homology group $H_m(C_{m,\mathbf{z}}, \mathcal{L}^*)$ of $C_{m,\mathbf{z}}$ with coefficients in the dual local system. This pairing is given by integration over cycles:

$$\omega \times \Delta \to \int_\Delta \omega, \qquad \omega \in H^m(C_{m,\mathbf{z}}, \mathcal{L}), \Delta \in H_m(C_{m,\mathbf{z}}, \mathcal{L}^*)$$

(about different choices of cycles and convergence of the corresponding integrals, cf. [51]). This pairing has the following property for any differential form $\omega$ on $C_{m,\mathbf{z}}$:

(6.10) $$\int_\Delta d \cdot \ell \omega = 0, \qquad \forall \Delta \in H_m(C_{m,\mathbf{z}}, \mathcal{L}^*).$$

We can now reproduce Schechtman-Varchenko integral formula.

**Theorem 4.** *Let $\Delta$ be an $m$–dimensional cycle on $C_{m,\mathbf{z}}$ with coefficients in the dual local system $\mathcal{L}^*$. The $\otimes_{i=1}^N M_{\lambda_i}$–valued function*

$$\prod_{i<j}(z_i - z_j)^{(\lambda_i,\lambda_j)/(k+h^\vee)} \int_\Delta \ell \, |w_1^{i_1}, \ldots, w_m^{i_m}\rangle \, dw_1 \ldots dw_m,$$

*where $|w_1^{i_1}, \ldots, w_m^{i_m}\rangle$ is the Bethe vector, is a solution of the KZ equation.*

*Proof.* To avoid confusion, we will denote by $A^{(z_i)}$ or $A^{(w_j)}$ the operator, which acts as $A$ on the factor $W_{\lambda_i,k}$ or $W_{-\alpha_{i_j},k}$ of the tensor product $\otimes_{i=1}^N W_{\lambda_i,k} \otimes_{j=1}^m W_{-\alpha_{i_j},k}$, and as the identity on all other factors.

From formula (6.7) we obtain:

$$\frac{\partial}{\partial z_i} \tau_{N,m}^k(\mathbf{z}, \mathbf{w}) = L_{-1}^{(z_i)*} \cdot \tau_{N,m}^k(\mathbf{z}, \mathbf{w}).$$

By restricting this equation, we have:

(6.11) $$\frac{\partial}{\partial z_i} \tau_{N,m}^k(\mathbf{z}, \mathbf{w})(\omega, G_{i_1}(-1)v_{-\alpha_{i_1}}, \ldots, G_{i_m}(-1)v_{-\alpha_{i_m}}) =$$
$$\tau_{N,m}^k(\mathbf{z}, \mathbf{w})(L_{-1}^{(z_i)} \cdot \omega, G_{i_1}(-1)v_{-\alpha_{i_1}}, \ldots, G_{i_m}(-1)v_{-\alpha_{i_m}}),$$

where $\omega \in \otimes_{i=1}^N M_{\lambda_i}^*$.

Using the Ward identity, we can rewrite the right hand side of equation (6.11) as

$$\frac{1}{k+h^\vee} \sum_{a=1}^d \sum_{j=1}^N \frac{1}{z_i - z_j} \tau_{N,m}^k(\mathbf{z}, \mathbf{w})(I_a^{(z_i)} I^{a(z_j)} \cdot \omega, G_{i_1}(-1)v_{-\alpha_{i_1}}, \ldots, G_{i_m}(-1)v_{-\alpha_{i_m}})$$

(6.12) $$+ \frac{1}{k+h^\vee} \sum_{a=1}^d \sum_{s=1}^m \sum_{n=0}^\infty \frac{1}{(z_i - w_s)^{n+1}} \tau_{N,m}^k(\mathbf{z}, \mathbf{w})(I_a^{(z_i)} \omega, G_{i_1}(-1)v_{-\alpha_{i_1}}, \ldots,$$
$$I^a(n) \cdot G_{i_s}(-1)v_{-\alpha_{i_s}}, \ldots, G_{i_m}(-1)v_{-\alpha_{i_m}}).$$

To compute the last term of this formula, we have to compute $I^a(n) \cdot G_{i_s}(-1) v_{-\alpha_{i_s}} \in W_{-\alpha_{i_s}, k}$ for $n \geq 0$.

**Lemma 5.** *For any $a = 1, \ldots, d$, and $i = 1, \ldots, l$,*

$$I^a(n) \cdot G_i(-1) v_{-\alpha_i} = 0, \qquad n > 1$$

*and there exists such $Y_a \in W_{-\alpha_i, k}$ that*

$$I^a(1) \cdot G_i(-1) v_{-\alpha_i} = (k + h^\vee) Y_a, \qquad I^a(0) \cdot G_i(-1) v_{-\alpha_i} = (k + h^\vee) L_{-1} Y_a.$$

This Lemma will be proved in the Appendix. It gives for the $s$th summand of the second term of formula (6.12):

$$(6.13) \qquad \tau_{N,m}^k(\mathbf{z}, \mathbf{w}) \left( \sum_{a=1}^d I_a^{(z_i)} \omega, G_{i_1}(-1) v_{-\alpha_{i_1}}, \ldots, \underset{s}{Y_a^s}, \ldots, G_{i_m}(-1) v_{-\alpha_{i_m}} \right),$$

where

$$Y_a^s = \frac{Y_a}{(z_i - w_s)^2} + \frac{L_{-1} Y_a}{z_i - w_s}.$$

According to (6.7), the action of $L_{-1}^{(w_s)*}$ on $\tau_{N,m}^k(\mathbf{z}, \mathbf{w})$ coincides with the action of $\partial/\partial w_s$. Therefore we can replace $Y_a^s$ by

$$\frac{\partial}{\partial w_s} \cdot \frac{Y_a}{z_i - w_s}.$$

Then formula (6.13) can be rewritten as

$$\frac{\partial}{\partial w_s} \left[ \frac{1}{z_i - w_s} \tau_{N,m}^k(\mathbf{z}, \mathbf{w}) \left( \sum_{a=1}^d I_a^{(z_i)} \omega, G_{i_1}(-1) v_{-\alpha_{i_1}}, \ldots, \underset{s}{Y_a}, \ldots, G_{i_m}(-1) v_{-\alpha_{i_m}} \right) \right].$$

For any $\mathbf{z}$ and $\omega$, the term in brackets, $V$, is a product of the multivalued function $\ell$ and a rational function on $C_{m,\mathbf{z}}$. Consider $V \, dw_1 \ldots w_m$ as an element of $\Omega^m(\mathcal{L})$. Then we have

$$\left( \frac{\partial}{\partial w_s} V \right) dw_1 \ldots dw_m = d(V \, dw_1 \ldots \widehat{dw_s} \ldots dw_m).$$

Therefore if $\Delta$ is an $m$–dimensional cycle on $C_{m,\mathbf{z}}$ with coefficients in $\mathcal{L}^*$, we have, by formula (6.10):

$$\int_\Delta \left( \frac{\partial}{\partial w_s} V \right) dw_1 \ldots dw_m = 0.$$

Thus we see that the second term of formula (6.12) will disappear after integration over $\Delta$.

But the first term coincides with the action of the operator $\Xi_i/(k+h^\vee)$. So, we obtain

$$\int_\Delta \tau_{N,m}^k(\mathbf{z},\mathbf{w})(L_{-1}^{(z_i)}\cdot\omega, G_{i_1}(-1)v_{-\alpha_{i_1}},\ldots,G_{i_m}(-1)v_{-\alpha_{i_m}})\,dw_1\ldots dw_m =$$

$$= \int_\Delta \frac{\Xi_i}{k+h^\vee}\tau_{N,m}^k(\mathbf{z},\mathbf{w})(\omega, G_{i_1}(-1)v_{-\alpha_{i_1}},\ldots,G_{i_m}(-1)v_{-\alpha_{i_m}})\,dw_1\ldots dw_m.$$

Hence, by (6.11),

$$(k+h^\vee)\frac{\partial}{\partial z_i}\int_\Delta \tau_{N,m}^k(\mathbf{z},\mathbf{w})(\omega, G_{i_1}(-1)v_{-\alpha_{i_1}},\ldots,G_{i_m}(-1)v_{-\alpha_{i_m}})\,dw_1\ldots dw_m =$$

$$= \Xi_i\int_\Delta \tau_{N,m}^k(\mathbf{z},\mathbf{w})(\omega, G_{i_1}(-1)v_{-\alpha_{i_1}},\ldots,G_{i_m}(-1)v_{-\alpha_{i_m}})\,dw_1\ldots dw_m,$$

for an arbitrary $\omega\in\otimes_{i=1}^N M_{\lambda_i}^*$.

The last formula can be rewritten as

$$(k+h^\vee)\frac{\partial}{\partial z_i}\int_\Delta \Psi^k(\mathbf{z},w_1^{i_1},\ldots,w_m^{i_m})\,dw_1\ldots dw_m = \Xi_i\int_\Delta \Psi^k(\mathbf{z},w_1^{i_1},\ldots,w_m^{i_m})\,dw_1\ldots dw_m.$$

Therefore we obtain

$$(k+h^\vee)\frac{\partial}{\partial z_i}\int_\Delta \Phi^k(\mathbf{z},w_1^{i_1},\ldots,w_m^{i_m})\,dw_1\ldots dw_m = \Xi_i\int_\Delta \Phi^k(\mathbf{z},w_1^{i_1},\ldots,w_m^{i_m})\,dw_1\ldots dw_m.$$

Theorem now follows from formula (6.9). $\square$

Theorem 4 shows that Bethe vectors enter solutions of the KZ equation. This has been observed in [36, 37, 38, 39]. Wakimoto modules provide a tool for constructing eigenvectors in Gaudin's model and for solving the KZ equation, thus giving a natural interpretation of this phenomenon.

## 7. Appendix.

In this Appendix we give the proof of Lemma 5 and Lemma 2, following [11], Appendix B.

First we prove Lemma 5 in the case $\mathfrak{g}=\mathfrak{sl}_2$. This is a direct computation, using the explicit formulas for the homomorphism $\rho$ (and hence $\rho_k$) given in Sect. 4.

We have $G(z)=-a(z)=-E(z)$. From the commutation relations of $\widehat{\mathfrak{sl}}_2$ we obtain

$$[E(n),G(-1)]=0,\quad [H(n),G(-1)]=2G(n-1),\quad [F(n),G(-1)]=H(n-1)-k\delta_{n,1}.$$

Since

$$E(n)v_{-\alpha}=G(n)v_{-\alpha}=0, n\geq 0,\qquad H(n)v_{-\alpha}=F(n)v_{-\alpha}=0, n>0,$$

and $H(0)v_{-\alpha}=-2v_{-\alpha}$, we conclude that

$$E(n)\cdot G(-1)v_{-\alpha}=H(n)\cdot G(-1)v_{-\alpha}=0,\qquad n\geq 0,$$

and
$$F(n) \cdot G(-1)v_{-\alpha} = 0, \qquad n > 1.$$
We also have:
$$F(1) \cdot G(-1)v_{-\alpha} = -(k+2)v_{-\alpha},$$
and
$$F(0) \cdot G(-1)v_{-\alpha} = G(-1)F(0)v_{-\alpha} + H(0)v_{-\alpha} =$$
$$= (2a(-1)a^*(0) - 2a(-1)a^*(0) + b(-1))v_{-\alpha} = -(k+2)L_{-1}v_{-\alpha}.$$
This proves Lemma 5 in the case $\mathfrak{g} = \mathfrak{sl}_2$.

Recall that in [40], Sect. 5.2, generalized Wakimoto modules over $\widehat{\mathfrak{g}}$, associated to an arbitrary parabolic subalgebra of $\mathfrak{g}$, were defined. Denote by $\mathfrak{p}^i$ the parabolic subalgebra of $\mathfrak{g}$, which is obtained by adjoining the generator $F_i$ of $\mathfrak{g}$ to the Borel subalgebra of $\mathfrak{g}$. Denote by $\mathfrak{sl}_2^i$ the Lie subalgebra of $\mathfrak{g}$ generated by $E_i, H_i$, and $F_i$. We have an orthogonal decomposition of $\mathfrak{p}^i$ into its semi-simple part $\mathfrak{sl}_2^i$ and abelian part $\mathfrak{h}_i^\perp$.

Denote by $\Gamma^i$ the Heisenberg Lie algebra with generators $a_\alpha(n), a_\alpha^*(n), \alpha \neq \alpha_i, n \in \mathbb{Z}$, and the same commutation relations as in $\Gamma(\mathfrak{g})$, cf. Sect. 4. Denote by $M^i$ the Fock representation of $\Gamma^i$, which is generated by a vacuum vector $v$, satisfying conditions (4.1).

The Wakimoto modules over $\widehat{\mathfrak{g}}$ of level $k$ associated to $\mathfrak{p}^i$ are realized in the tensor product $M^i \otimes \pi_\chi^i \otimes L$. Here $\pi_\chi^i$ is the Fock representation of the Heisenberg Lie algebra associated to $\mathfrak{h}_i^\perp$ and $L$ is a representation of $\widehat{\mathfrak{sl}}_2$ of level $k'$ such that

$$(7.1) \qquad k + h^\vee = \frac{(\alpha_i, \alpha_i)}{2}(k' + 2).$$

*Remark* 6. We would like to take this opportunity to correct some errors in [40], Sect. 5.2.

First, in the exact sequences on page 179 the module $End_p$ should be replaced by $A_p^0$.

Second, in formula (4) on page 180 one should take into account the difference in normalization of the invariant scalar product $\langle \cdot, \cdot \rangle$ for a Lie algebra and its subalgebra. Let $\delta_i$ be the ratio between the scalar product on the Lie algebra $g$ and its Lie subalgebra $g_p^{(i)}$ (we use the notation from [40]). Then formula (4) should be rewritten as

$$k_p^{(i)} + c_{g_p^{(i)}} = \delta_i k_p'.$$

Since $k_p' = k + c_g$, this gives the following condition on levels of representations of $g_p^{(i)}$ and $g$:

$$k_p^{(i)} + c_{g_p^{(i)}} = \delta_i(k + c_g).$$

In particular, in our case $p = \mathfrak{p}^i$, so $\delta = 2/(\alpha_i, \alpha_i)$, and we obtain formula (7.1). □

Let us choose as $L$ over $\mathfrak{sl}_2^i$, the module $W_{y \cdot \alpha_i, k'}, y \in \mathbb{C}$. This is a module over the Heisenberg algebra generated by $a_{\alpha_i}(n), a^*_{\alpha_i}(n), b_i(n), n \in \mathbb{Z}$. It is clear that the resulting module over $\widehat{\mathfrak{g}}$ is isomorphic to the standard Wakimoto module $W_{y \cdot \alpha_i, k}$, on which the operators $G(n)$ act as $-a_{-\alpha_i}(n)$. In other words, we choose such coordinates $x_\alpha$ on the big cell $U$ of the flag manifold of $\mathfrak{g}$ in which $G_i = -\partial/\partial x_{\alpha_i}$. In particular, we see that the action of the operators $E_i(n), H_i(n)$, and $F_i(n)$ with $n \geq 0$ on the subspace $W_{y \cdot \alpha_i, k'} \subset W_{y \cdot \alpha_i, k}$ is the same as the action of the operators $E(n), H(n)$, and $F(n)$ on the $\mathfrak{sl}_2$-module $W_{y \cdot \alpha_i, k'}$.

Hence the action of the operators $E_i(n), H_i(n)$, and $F_i(n)$ on $G_i(-1)v_{-\alpha_i}$ coincides with that in the case of $\mathfrak{sl}_2$. Thus, we obtain:

$$E_i(n) \cdot G_i(-1)v_{-\alpha_i} = H_i(n) \cdot G_i(-1)v_{-\alpha_i} = 0, \qquad n \geq 0,$$

$$F_i(n) \cdot G_i(-1)v_{-\alpha_i} = 0, \qquad n > 1,$$

and

$$(7.2) \qquad F_i(1) \cdot G_i(-1)v_{-\alpha_i} = -\frac{2}{(\alpha_i, \alpha_i)}(k + h^\vee)v_{-\alpha_i},$$

$$(7.3) \qquad F_i(0) \cdot G_i(-1)v_{-\alpha_i} = -\frac{2}{(\alpha_i, \alpha_i)}(k + h^\vee)L_{-1}v_{-\alpha_i}.$$

Now consider the operators $E_j(n), H_j(n)$, and $F_j(n)$ with $j \neq i$. We first compute the finite-dimensional commutation relations of the differential operators $\bar{\rho}_\chi[E_j], \bar{\rho}_\chi[H_j]$, and $\bar{\rho}_\chi[F_j]$, with the vector field $G_i$ (cf. Sect. 4). These commutation relations read (cf., e.g., [50], Sect. 3):

$$[\bar{\rho}_\chi[E_j], G_i] = 0, \qquad [\bar{\rho}_\chi[H_j], G_i] = \alpha_i(H_j)G_i,$$

and

$$[\bar{\rho}_\chi[F_j], G_i] = \alpha_i(H_j)x_{\alpha_j}G_i.$$

Using these formulas, we can find the commutation relations of $E_j(n), H_j(n), F_j(n)$, $j \neq i$, with $G_i(m)$, using the Wick theorem. Since $G_i(z) = -a_{\alpha_i}(z)$, there can be no "double contractions", and there is no contribution from the "coboundary term" $c_j \partial a^*_{\alpha_j}(z)$. Therefore this computation amounts to the computation of all "single contractions". But those are uniquely defined by the finite-dimensional commutation relations above. So we obtain:

$$[E_j(n), G_i(z)] = 0, \qquad [H_j(n), G_i(z)] = z^{-n}\alpha_i(H_j)G_i(z),$$

and

$$[F_j(n), G_i(z)] = z^{-n}\alpha_i(H_j) : a^*_{\alpha_j}(z)G_i(z) : .$$

These formulas immediately give:

$$E_j(n) \cdot G_i(-1)v_{-\alpha_i} = G_i(-1)E_j(n)v_{-\alpha_i} = 0, \quad n \geq 0,$$

$$H_j(n) \cdot G_i(-1)v_{-\alpha_i} = G_i(-1)H_j(n)v_{-\alpha_i} + \alpha_i(H_j)G_i(n-1)v_{-\alpha_i} = 0,$$

$$F_j(n) \cdot G_i(-1)v_{-\alpha_i} = G_i(-1)F_j(n)v_{-\alpha_i} + \alpha_i(H_j) \sum_{s+r=n-1} a^*_{\alpha_j}(r)G_i(s)v_{-\alpha_i} = 0.$$

The last two formulas clearly hold for $n > 0$, and also for $n = 0$, because $H_j(0)v_{-\alpha_i} = -\alpha_i(H_j)v_{-\alpha_i}$ and $F_j(0)v_{-\alpha_i} = -\alpha_i(H_j)a^*_{\alpha_j}(0)v_{-\alpha_i}$.

We can compute the action of any other element of $\widehat{\mathfrak{g}}$ of the form $I^a(n), n \geq 0$, on $G_i(-1)v_{-\alpha_i}$, using its presentation as a commutator of $E_i(m)$'s or $F_i(m)$'s, depending on whether $I^a$ belongs to the upper or lower nilpotent subalgebra of $\mathfrak{g}$, respectively (if $I^a \in \mathfrak{h}$, then it is a linear combinations of $H_i$'s and there is nothing to prove). Clearly, in the first case we obtain 0. In the second case, we can realize $I^a(n)$ as a successive commutator of $E_i(n), n \geq 0$, and $E_j(0), j \neq i$. The statement of Lemma 5 then follows from formulas (7.2), (7.3), and the relation $[L_{-1}, A(0)] = 0$ for any $A \in \mathfrak{g}$.

Lemma 2 can be proved along the same lines. We find:

$$E_j(n) \cdot G_i(-1)v = H_j(n) \cdot G_i(-1)v = 0,$$

and

$$F_j(n) \cdot G_i(-1)v = 0, n > 0, \qquad F_j(0) \cdot G_i(-1)v = \delta_{i,j}\mu^{(0)}(H_i)v$$

for any $j = 1, \ldots, l$ and $n \geq 0$. Therefore $G_i(-1)$ is a singular vector of imaginary degree, if and only if $\mu^{(0)}(H_i) = 2(\mu^{(0)}, \alpha_i)/(\alpha_i, \alpha_i) = 0$.

**Acknowledgements.** We thank A. Beilinson, V. Ginzburg, D. Kazhdan, and E. Sklyanin for useful discussions.

The research of E. F. was supported by a Junior Fellowship from the Society of Fellows of Harvard University and by NSF grant DMS-9205303. The research of N. R. was supported by Alfred P. Sloan fellowship and by NSF grant DMS-9296120.

LANDAU INSTITUTE FOR THEORETICAL PHYSICS, KOSYGINA ST 2, MOSCOW 117940, RUSSIA

DEPARTMENT OF MATHEMATICS, HARVARD UNIVERSITY, CAMBRIDGE, MA 02138, USA

DEPARTMENT OF MATHEMATICS, UNIVERSITY OF CALIFORNIA, BERKELEY, CA 94720, USA